\documentclass[12pt, draftclsnofoot, onecolumn]{IEEEtran}

\usepackage{ifthen} 

\newboolean{useDraftClass}
\setboolean{useDraftClass}{true} 

\ifthenelse{\boolean{useDraftClass}}{
    \usepackage{setspace}
    \onecolumn
}{
}

\usepackage{float}
\usepackage{tikz,pgfplots}
\usetikzlibrary{arrows}

\usepackage[utf8]{inputenc} 
\usepackage[T1]{fontenc}
\usepackage{url}
\usepackage{ifthen}

\usepackage{cite}

\usepackage[cmex10]{amsmath} 
\usepackage{amsmath,amssymb,amsfonts} 
\usepackage{graphicx}
\usepackage{xcolor}

\usepackage{cuted}
\usepackage{comment}

\usepackage{enumitem}
\setlist[itemize]{leftmargin=*}

\usepackage{physics}
\usepackage{amsmath}
\usepackage{tikz}
\usepackage{mathdots}
\usepackage{yhmath}
\usepackage{cancel}
\usepackage{color}
\usepackage{siunitx}
\usepackage{array}
\usepackage{multirow}
\usepackage{amssymb}
\usepackage{gensymb}
\usepackage{tabularx}
\usepackage{extarrows}
\usepackage{booktabs}
\usetikzlibrary{fadings}
\usetikzlibrary{patterns}
\usetikzlibrary{shadows.blur}
\usetikzlibrary{shapes}

\usepackage{adjustbox}
\usepackage{tikz,pgfplots}

\usepackage[hidelinks]{hyperref}  

\usepackage{environ}         
\usepackage{etoolbox}        
\usepackage{graphicx}        

\usepackage{caption}
\usepackage{subcaption}

\usepackage{authblk}

\usepackage{amsthm}
\newtheorem{theorem}{Theorem}
\newtheorem{lemma}[theorem]{Lemma}

\newtheorem{corollary}{Corollary}

\allowdisplaybreaks

\newlength{\myl}
\let\origequation=\equation
\let\origendequation=\endequation

\RenewEnviron{equation}{
	\settowidth{\myl}{$\BODY$}                       
	\origequation
	\ifdimcomp{\the\linewidth}{>}{\the\myl}
	{\ensuremath{\BODY}}                             
	{\resizebox{\linewidth}{!}{\ensuremath{\BODY}}}  
	\origendequation
}

\begin{document}
\title{Capacity Approximations for Insertion Channels with Small Insertion Probabilities} 

\author[1]{Busra Tegin}
\author[2]{Tolga M. Duman\thanks{
This work was funded by the European Union through the ERC Advanced Grant 101054904: TRANCIDS. Views and opinions expressed are, however, those of the authors only and do not necessarily reflect those of the European Union or the European Research Council Executive Agency. Neither the European Union nor the granting authority can be held responsible for them. \\
A preliminary version of this work was presented at the 2025 IEEE International Symposium on Information Theory (ISIT),
Ann Arbor, MI, USA, June 2025 [DOI: 10.1109/ISIT63088.2025.11195329] \\
\textit{Part of this work was conducted while Busra Tegin was affiliated with Bilkent University.}
}}

\affil[1]{\small IETR - UMR CNRS 6164, CentraleSupélec, avenue de la Boulaie - CS 47601 35576\protect\\CESSON-SEVIGNE Cedex, France}

\affil[2]{\small Department of Electrical and Electronics Engineering\protect\\Bilkent University, Ankara, Turkey}

\affil[ ]{Email: {\tt  busra.tegin@centralesupelec.fr,~duman@ee.bilkent.edu.tr}}

\maketitle 

\begin{abstract}
Channels with synchronization errors, exhibiting deletion and insertion errors, find practical applications in DNA storage, data reconstruction, and various other domains. The presence of insertions and deletions renders the channel with memory, complicating capacity analysis. For instance, despite the formulation of an independent and identically distributed (i.i.d.) deletion channel more than fifty years ago, and proof that the channel is information stable, hence its Shannon capacity exists, calculation of the capacity remained elusive. However, a relatively recent result establishes the capacity of the deletion channel in the asymptotic regime of small deletion probabilities by computing the dominant terms of its capacity expansion. This paper extends that result to binary insertion channels, determining the dominant terms of the channel capacity for small insertion probabilities and establishing capacity in this asymptotic regime. Specifically, we consider two i.i.d. insertion channel models: the simple insertion channel, where a random bit may be inserted after each transmitted bit, and the Gallager insertion model, for which a bit is replaced by two random bits with a certain probability. To prove our results, we build on methods used for the deletion channel, employing Bernoulli$(1/2)$ inputs for achievability and coupling this with a converse using stationary and ergodic input processes, and show that the channel capacity differs only in the higher order terms from the achievable rates with i.i.d. inputs. The results, for instance, show that the capacity of the simple insertion channel is higher than that of the Gallager insertion channel, and quantify the difference in the asymptotic regime.
\end{abstract}


\section{Introduction}

Synchronization-error channels, including deletion, repetition, and insertion channels, serve as effective models for a variety of realistic and practical scenarios. Such channel models find extensive applications in DNA coding \cite{abroshan2019coding, balado2010embedding}, DNA-based storage systems \cite{heckel2019characterization, lenz2019coding}, magnetic data storage systems \cite{guan2014coding}, data reconstruction \cite{levenshtein2001efficient}, and numerous other domains. Despite the wide range of application areas, the capacity of channels with synchronization errors remains an underexplored research area. In \cite{dobrushin1967shannon}, Dobrushin establishes counterparts to Shannon's discrete memoryless channel's (DMC's) capacity formulation \cite{shannon} for a wide range of channels with synchronization errors. He demonstrates that the channel capacity for such channels can be defined by the maximal mutual information per bit and can be achieved through a random coding scheme. 
In \cite{10619598, morozov2024shannon}, the authors consider channels with synchronization errors governed by a Markov chain. It is shown that if the underlying Markov chain is stationary and ergodic, the information capacity exists and is equal to the coding capacity. This result generalizes Dobrushin's coding theorem \cite{dobrushin1967shannon} on the information stability of channels with memoryless synchronization errors. 

Despite the early result establishing the information stability of the memoryless synchronization error channels, the calculation of the exact capacity proved extremely challenging. For instance, even for the seemingly simple channel model of independent and identically distributed (i.i.d.) deletion channel, the channel capacity is unknown. As such the literature on channels with synchronization errors predominantly focuses on establishing lower and upper bounds for channel capacity. In \cite{gallager2000sequential}, Gallager proposed using a pseudo-random string to convolutional codes with sequential decoding to correct synchronization and substitution errors. Tree codes are utilized in \cite{zigangirov1969sequential} to obtain lower bounds for insertion/deletion channels. In \cite{mitzenmacher2006simple}, a lower bound for the deletion channel is established by demonstrating a connection between deletion channels and the Poisson-repeat channel using jig-saw decoding. This lower bound is further improved in \cite{kirsch2009directly} for the capacity of channels with independent and identically distributed (i.i.d.) deletions and duplications. 

To develop upper bounds on the capacity of channels with synchronization errors, typical technique is to employ a genie-aided approach with additional side information, which results in more tractable channel models. In \cite{diggavi2007capacity}, an upper bound on the deletion channel capacity is derived using undeletable markers. Similarly, Ref. \cite{fertonani2010novel} provides deletion channel upper bounds coupled with lower bounds, by using different genie-aided systems. In \cite{rahmati2013upper}, an upper bound for the non-binary deletion channel is derived. Furthermore, in \cite{fertonani2010bounds}, capacity upper bounds are obtained for more general synchronization error channels exhibiting insertions, deletions and substitutions. 
In \cite{10.1145/3281275}, using a systematic approach based on convex programming and real analysis, the author derives upper bounds on the capacity of the binary deletion channel, channels with i.i.d. insertions and deletions, and the Poisson-repeat channel. This framework can be applied to obtain capacity upper bounds for any repetition distribution.
A recent and comprehensive overview of the capacity results of synchronization-error channels is provided in \cite{cheraghchi2020overview}. 

Parallel to the results establishing upper and lower bounds on the channel capacity, there have also been efforts determining the behavior of the capacity of the deletion channel in the asymptotic regime of small deletion probabilities. For instance, \cite{5513746} demonstrates that the capacity of the deletion channel is comparable to that of the binary symmetric error channel in the asymptotic case for small deletion probabilities. The authors establish that the trivial capacity lower bound for this channel, given by $1 - H(p)$, is tight in the limit. They provide an upper bound of $1 - (1 - o(1))H(p)$, where the $o(1)$ term vanishes as the deletion probability $p \rightarrow 0$. More in line with our work in this paper, the works presented in \cite{kanoria2010deletion, kanoria2013optimal} leverage Dobrushin’s coding theorem \cite{dobrushin1967shannon} to approximate the capacity of the binary deletion channel using series expansions up to the second and third leading terms, respectively, for small deletion probabilities. These studies establish the channel’s capacity in the asymptotic regime of small deletion probabilities by calculating the dominant terms of the capacity expansion. Additionally, they demonstrate that the optimal input distribution can be obtained by perturbing the i.i.d. Bernoulli$(1/2)$ input distribution.
In \cite{6457365}, a similar approach is applied to duplication channels, where the channel capacity is computed for small values of the duplication probability using a series expansion. Note that although the duplication channel is closely related to the insertion channel and shares similarities with our channel model, it duplicates existing bits/symbols rather than inserting random bits/symbols, hence no runs are lost and no new runs are created, significantly simplifying the capacity calculation. Indeed, the capacity of the duplication channel can be computed with arbitrary {\color{black}{precision}} for any duplication probability \cite{4418490}.

In this paper, we adopt a systematic approach similar to that of~\cite{kanoria2010deletion, kanoria2013optimal}, but applied to insertion channels, which serve as practical models for real-world communication scenarios such as DNA storage, where insertions are a prevalent type of error. We focus on two fundamental binary insertion channel models: 1) the simple insertion channel, in which each transmitted bit is followed by a random bit with a certain small probability, and 2) the Gallager insertion channel, in which each bit is replaced by two random bits with a certain small probability. In both cases, we focus on small insertion probabilities. 
We derive the first two terms in the asymptotic expansion of the channel capacities and establish their achievability using an i.i.d. Bernoulli$(1/2)$ input distribution. We also provide a matching converse for the dominant (lower-order) term, thereby obtaining an asymptotic approximation to the channel capacity, rather than a lower or upper bound.

While our analytical strategy draws inspiration from the techniques developed for the deletion channel \cite{kanoria2010deletion, kanoria2013optimal}, the insertion channel presents fundamentally different challenges. Notably, the statistical behavior of insertion errors complicates the alignment between input and output sequences in a manner distinct from deletions. As a result, the entropy calculations, derivation of bounds, and design of auxiliary distributions require novel adaptations, even when the high-level proof structure resembles prior work. As the first work that provides capacity approximations for the insertion channels, the results provide a significant step toward understanding the capacity of synchronization channels beyond deletions.

Our main results are as follows: let $C(\alpha)$ be the capacity of an insertion channel with insertion probability $\alpha$. Then, for the simple insertion channel model with small $\alpha$ and any $\epsilon>0$, we have
\begin{align}
    C_1(\alpha) = 1 + \alpha \log(\alpha)  +G_1\alpha + \mathcal{O}(\alpha^{3/2 - \epsilon}),
\end{align}
where 
\begin{align*}
G_1 = - &\log(e) + \frac{1}{2}\sum_{l=1}^{\infty} 2^{-l-1}l \log l +  \frac{1}{2}\sum_{a = 1}^\infty  (a+1)2^{-a}h\left(\frac{1}{a+1} \right).
\end{align*}
For the Gallager insertion channel, we obtain 
\begin{align}
    C_2(\alpha) = 1 + \alpha \log(\alpha)  +G_2\alpha + \mathcal{O}(\alpha^{3/2 - \epsilon}),
\end{align}
with
 \begin{align*}
     G_2 = - &\log(e) - \frac{7}{8}
     +\frac{1}{4} \sum_{l=1}^{\infty} 2^{-l-1}l \log l  + \frac{1}{4} \sum_{a = 1}^\infty \sum_{b = 1}^\infty  (a+b+2)2^{-a}2^{-b}h\left(\frac{a+1}{a+b+2} \right),
 \end{align*}
where $G_1 \approx 0.4901$, and $G_2 \approx  -0.5865$, respectively. 

\begin{figure}[h]
        \centering
%
%
\definecolor{mycolor1}{rgb}{0.63529,0.07843,0.18431}%
\definecolor{mycolor2}{rgb}{0.00000,0.44706,0.74118}%
\definecolor{mycolor3}{rgb}{0.00000,0.49804,0.00000}%
\definecolor{mycolor4}{rgb}{0.87059,0.49020,0.00000}%
\definecolor{mycolor5}{rgb}{0.00000,0.44700,0.74100}%
\definecolor{mycolor6}{rgb}{0.74902,0.00000,0.74902}%
\begin{tikzpicture}

\begin{axis}[%
width=12cm,
height=5cm,
at={(0.745in,0.444in)},
scale only axis,
unbounded coords=jump,
xmin=0,
xmax=0.25,
xlabel style={font=\color{white!15!black}},
xlabel={$\alpha\text{ (insertion probability)}$},
ymin=0,
ymax=1,
ylabel style={font=\color{white!15!black}},
ylabel={Capacity approximation},
axis background/.style={fill=white},
xmajorgrids,
ymajorgrids,
legend style={at={(0.375,0.25)},font=\small,nodes={scale=0.85, transform shape},
legend cell align={left},
        /tikz/column 2/.style={
            column sep=5pt,
        },
    },
xtick={0.05,0.1, 0.15, 0.2, 0.25},
xticklabel style={/pgf/number format/fixed},
]
\addplot [color=mycolor1, line width=1.5pt, mark=x, mark options={solid, mycolor1}]
  table[row sep=crcr]{%
0	1\\
0.01	0.938462456830054\\
0.02	0.896924913660108\\
0.03	0.862936245511796\\
0.04	0.833849827320216\\
0.05	0.808408688894638\\
0.06	0.785872491023593\\
0.07	0.76575204235441\\
0.08	0.747699654640431\\
0.09	0.731455361600293\\
0.1	0.716817377789275\\
0.11	0.703624503180695\\
0.12	0.691744982047185\\
0.13	0.681069102149043\\
0.14	0.671504084708819\\
0.15	0.662970441792086\\
0.16	0.655399309280862\\
0.17	0.648730449123474\\
0.18	0.642910723200586\\
0.19	0.637892907325305\\
0.2	0.63363475557855\\
0.21	0.630098252214671\\
0.22	0.627249006361391\\
0.23	0.625055756984352\\
0.24	0.623489964094371\\
0.25	0.622525468195028\\
0.26	0.622138204298085\\
0.27	0.622305959995592\\
0.28	0.623008169417638\\
0.29	0.624225736658559\\
0.3	0.625940883584172\\
0.31	0.628137017951602\\
0.32	0.630798618561724\\
0.33	0.633911134780068\\
0.34	0.637460898246948\\
0.35	0.641435044982624\\
0.36	0.645821446401173\\
0.37	0.650608647994705\\
0.38	0.65578581465061\\
0.39	0.661342681728379\\
0.4	0.667269511157101\\
0.41	0.673557051925624\\
0.42	0.680196504429343\\
0.43	0.687179488214219\\
0.44	0.694498012722782\\
0.45	0.702144450700779\\
0.46	0.710111513968705\\
0.47	0.718392231301022\\
0.48	0.726979928188742\\
0.49	0.735868208289093\\
0.5	0.745050936390057\\
0.51	0.754522222738211\\
0.52	0.76427640859617\\
0.53	0.774308052911352\\
0.54	0.784611919991183\\
0.55	0.795182968091527\\
0.56	0.806016338835276\\
0.57	0.817107347386975\\
0.58	0.828451473317117\\
0.59	0.840044352096666\\
0.6	0.851881767168345\\
0.61	0.863959642546648\\
0.62	0.876274035903204\\
0.63	0.888821132098342\\
0.64	0.901597237123449\\
0.65	0.914598772421998\\
0.66	0.927822269560136\\
0.67	0.941264365220318\\
0.68	0.954921796493895\\
0.69	0.968791396450655\\
0.7	0.982870089965249\\
0.71	0.99715488978215\\
0.72	1.01164289280234\\
0.73	1.02633127657635\\
0.74	1.04121729598941\\
0.75	1.05629828012595\\
0.76	1.07157162930122\\
0.77	1.08703481224922\\
0.78	1.10268536345676\\
0.79	1.11852088063417\\
0.8	1.1345390223142\\
0.81	1.15073750557091\\
0.82	1.16711410385125\\
0.83	1.18366664491242\\
0.84	1.20039300885869\\
0.85	1.21729112627163\\
0.86	1.23435897642844\\
0.87	1.25159458560308\\
0.88	1.26899602544556\\
0.89	1.28656141143489\\
0.9	1.30428890140156\\
0.91	1.32217669411572\\
0.92	1.34022302793741\\
0.93	1.35842617952548\\
0.94	1.37678446260204\\
0.95	1.39529622676952\\
0.96	1.41395985637748\\
0.97	1.43277376943674\\
0.98	1.45173641657819\\
0.99	1.47084628005415\\
1	1.49010187278011\\
};
\addlegendentry{Simple insertion ch.}

\addplot [color=mycolor2, line width=1.5pt, mark=o, mark options={solid, mycolor2}]
  table[row sep=crcr]{%
0	1\\
0.01	0.927696247757362\\
0.02	0.875392495514723\\
0.03	0.83063761829372\\
0.04	0.790784991029447\\
0.05	0.754577643531177\\
0.06	0.721275236587439\\
0.07	0.690388578845564\\
0.08	0.661569982058893\\
0.09	0.634559479946063\\
0.1	0.609155287062353\\
0.11	0.585196203381081\\
0.12	0.562550473174879\\
0.13	0.541108384204044\\
0.14	0.520777157691128\\
0.15	0.501477305701703\\
0.16	0.483139964117787\\
0.17	0.465704894887706\\
0.18	0.449118959892126\\
0.19	0.433334934944153\\
0.2	0.418310574124706\\
0.21	0.404007861688135\\
0.22	0.390392406762162\\
0.23	0.377432948312431\\
0.24	0.365100946349758\\
0.25	0.353370241377723\\
0.26	0.342216768408087\\
0.27	0.331618315032902\\
0.28	0.321554315382256\\
0.29	0.312005673550484\\
0.3	0.302954611403406\\
0.31	0.294384536698143\\
0.32	0.286279928235573\\
0.33	0.278626235381225\\
0.34	0.271409789775412\\
0.35	0.264617727438397\\
0.36	0.258237919784253\\
0.37	0.252258912305093\\
0.38	0.246669869888306\\
0.39	0.241460527893382\\
0.4	0.236621148249412\\
0.41	0.232142479945243\\
0.42	0.22801572337627\\
0.43	0.224232498088454\\
0.44	0.220784813524324\\
0.45	0.217665042429629\\
0.46	0.214865896624863\\
0.47	0.212380404884488\\
0.48	0.210201892699515\\
0.49	0.208323963727174\\
0.5	0.206740482755446\\
0.51	0.205445560030908\\
0.52	0.204433536816175\\
0.53	0.203698972058664\\
0.54	0.203236630065803\\
0.55	0.203041469093455\\
0.56	0.203108630764512\\
0.57	0.203433430243518\\
0.58	0.204011347100969\\
0.59	0.204838016807825\\
0.6	0.205909222806811\\
0.61	0.207220889112423\\
0.62	0.208769073396286\\
0.63	0.210549960518733\\
0.64	0.212559856471147\\
0.65	0.214795182697004\\
0.66	0.21725247076245\\
0.67	0.21992835734994\\
0.68	0.222819579550825\\
0.69	0.225922970434892\\
0.7	0.229235454876794\\
0.71	0.232754045621003\\
0.72	0.236475839568505\\
0.73	0.240398014269815\\
0.74	0.244517824610187\\
0.75	0.248832599674036\\
0.76	0.253339739776612\\
0.77	0.258036713651923\\
0.78	0.262921055786764\\
0.79	0.267990363891483\\
0.8	0.273242296498824\\
0.81	0.278674570682842\\
0.82	0.284284959890486\\
0.83	0.290071291878966\\
0.84	0.296031446752539\\
0.85	0.302163355092789\\
0.86	0.308464996176908\\
0.87	0.314934396278859\\
0.88	0.321569627048649\\
0.89	0.328368803965283\\
0.9	0.335330084859258\\
0.91	0.342451668500726\\
0.92	0.349731793249726\\
0.93	0.357168735765105\\
0.94	0.364760809768976\\
0.95	0.372506364863759\\
0.96	0.380403785399031\\
0.97	0.388451489385596\\
0.98	0.396647927454348\\
0.99	0.404991581857619\\
1	0.413480965510892\\
};
\addlegendentry{Gallager insertion ch.}

\end{axis}
\end{tikzpicture}%
        \caption{Capacity approximations for both insertion channels.}
        \label{fig:top_plot}
\end{figure}

\begin{figure}[h]
   
    \begin{subfigure}{\linewidth}
        \centering
        \input{figures/res_bound}
        \caption{Upper and lower bounds for the Gallager insertion channel model.}
        \label{fig:bottom_plot}
    \end{subfigure}

        \begin{subfigure}{\linewidth}
        \centering
%
%
\definecolor{mycolor1}{rgb}{0.63529,0.07843,0.18431}%
\definecolor{mycolor2}{rgb}{0.00000,0.44706,0.74118}%
\definecolor{mycolor3}{rgb}{0.00000,0.49804,0.00000}%
\definecolor{mycolor4}{rgb}{0.87059,0.49020,0.00000}%
\definecolor{mycolor5}{rgb}{0.00000,0.44700,0.74100}%
\definecolor{mycolor6}{rgb}{0.74902,0.00000,0.74902}%
\definecolor{mycolor7}{rgb}{0.47, 0.27, 0.23}
\begin{tikzpicture}

\begin{axis}[%
width=12cm,
height=5cm,
at={(0.745in,0.444in)},
scale only axis,
unbounded coords=jump,
xmin=0,
xmax=0.25,
xlabel style={font=\color{white!15!black}},
xlabel={$\alpha\text{ (insertion probability)}$},
ymin=0,
ymax=1,
ylabel style={font=\color{white!15!black}},
ylabel={Capacity approximation},
axis background/.style={fill=white},
xmajorgrids,
ymajorgrids,
legend style={at={(0.47,0.25)},font=\small,nodes={scale=0.85, transform shape},
legend cell align={left},
        /tikz/column 2/.style={
            column sep=5pt,
        },
    },
xtick={0.05,0.1, 0.15, 0.2, 0.25},
xticklabel style={/pgf/number format/fixed},
]

\addplot [color=mycolor7, dashdotted, mark=triangle, line width=1.5pt]
  table[row sep=crcr]{%
0	1\\
0.01	0.98102804352423\\
0.02	0.962685238339185\\
0.03	0.944955028977901\\
0.04	0.927821383516697\\
0.05	0.911268779893273\\
0.06	0.895282192505623\\
0.07	0.879847079088526\\
0.08	0.864949367865043\\
0.09	0.850575444971347\\
0.1	0.836712142154308\\
0.11	0.823346724742523\\
0.12	0.810466879893086\\
0.13	0.798060705118258\\
0.14	0.786116697098428\\
0.15	0.77462374079035\\
0.16	0.763571098842691\\
0.17	0.752948401334402\\
0.18	0.742745635855411\\
0.19	0.732953137953592\\
0.2	0.723561581976924\\
0.21	0.714561972345165\\
0.22	0.705945635291124\\
0.23	0.69770421111767\\
0.24	0.68982964702271\\
0.25	0.682314190550303\\
};
\addlegendentry{Capacity UB (Blahut-Arimoto)}

\addplot [color=mycolor1, line width=1.5pt, mark=x, mark options={solid, mycolor1}]
  table[row sep=crcr]{%
0	1\\
0.01	0.938462456830054\\
0.02	0.896924913660108\\
0.03	0.862936245511796\\
0.04	0.833849827320216\\
0.05	0.808408688894638\\
0.06	0.785872491023593\\
0.07	0.76575204235441\\
0.08	0.747699654640431\\
0.09	0.731455361600293\\
0.1	0.716817377789275\\
0.11	0.703624503180695\\
0.12	0.691744982047185\\
0.13	0.681069102149043\\
0.14	0.671504084708819\\
0.15	0.662970441792086\\
0.16	0.655399309280862\\
0.17	0.648730449123474\\
0.18	0.642910723200586\\
0.19	0.637892907325305\\
0.2	0.63363475557855\\
0.21	0.630098252214671\\
0.22	0.627249006361391\\
0.23	0.625055756984352\\
0.24	0.623489964094371\\
0.25	0.622525468195028\\
0.26	0.622138204298085\\
0.27	0.622305959995592\\
0.28	0.623008169417638\\
0.29	0.624225736658559\\
0.3	0.625940883584172\\
0.31	0.628137017951602\\
0.32	0.630798618561724\\
0.33	0.633911134780068\\
0.34	0.637460898246948\\
0.35	0.641435044982624\\
0.36	0.645821446401173\\
0.37	0.650608647994705\\
0.38	0.65578581465061\\
0.39	0.661342681728379\\
0.4	0.667269511157101\\
0.41	0.673557051925624\\
0.42	0.680196504429343\\
0.43	0.687179488214219\\
0.44	0.694498012722782\\
0.45	0.702144450700779\\
0.46	0.710111513968705\\
0.47	0.718392231301022\\
0.48	0.726979928188742\\
0.49	0.735868208289093\\
0.5	0.745050936390057\\
0.51	0.754522222738211\\
0.52	0.76427640859617\\
0.53	0.774308052911352\\
0.54	0.784611919991183\\
0.55	0.795182968091527\\
0.56	0.806016338835276\\
0.57	0.817107347386975\\
0.58	0.828451473317117\\
0.59	0.840044352096666\\
0.6	0.851881767168345\\
0.61	0.863959642546648\\
0.62	0.876274035903204\\
0.63	0.888821132098342\\
0.64	0.901597237123449\\
0.65	0.914598772421998\\
0.66	0.927822269560136\\
0.67	0.941264365220318\\
0.68	0.954921796493895\\
0.69	0.968791396450655\\
0.7	0.982870089965249\\
0.71	0.99715488978215\\
0.72	1.01164289280234\\
0.73	1.02633127657635\\
0.74	1.04121729598941\\
0.75	1.05629828012595\\
0.76	1.07157162930122\\
0.77	1.08703481224922\\
0.78	1.10268536345676\\
0.79	1.11852088063417\\
0.8	1.1345390223142\\
0.81	1.15073750557091\\
0.82	1.16711410385125\\
0.83	1.18366664491242\\
0.84	1.20039300885869\\
0.85	1.21729112627163\\
0.86	1.23435897642844\\
0.87	1.25159458560308\\
0.88	1.26899602544556\\
0.89	1.28656141143489\\
0.9	1.30428890140156\\
0.91	1.32217669411572\\
0.92	1.34022302793741\\
0.93	1.35842617952548\\
0.94	1.37678446260204\\
0.95	1.39529622676952\\
0.96	1.41395985637748\\
0.97	1.43277376943674\\
0.98	1.45173641657819\\
0.99	1.47084628005415\\
1	1.49010187278011\\
};
\addlegendentry{Simple insertion ch. (Cap. Approx.)}

\end{axis}
\end{tikzpicture}%
        \caption{Upper bound for the simple insertion channel model.}
        \label{fig:bottom_plot2}
    \end{subfigure}

    \caption{Bounds for the capacity of the insertion channels.}
    \label{fig:insertion_channels}
\end{figure}

Approximations obtained by omitting the terms with higher order for these channel models are illustrated in Fig. \ref{fig:top_plot}. As anticipated, the Gallager insertion channel replaces each bit with two random bits; hence, for a given input, the ambiguity in the channel output is higher compared to the simple insertion channel model with random bit insertion (without replacement). Therefore, intuitively, one would expect a lower rate for the Gallager insertion channel, which is verified by the results in Fig. \ref{fig:top_plot}. We note that these approximations are asymptotically exact as $\alpha \rightarrow 0$, and we expect that they remain as good approximations for small $\alpha$; however, there are no guarantees for any specific non-zero $\alpha$ value. 
Furthermore, we provide upper and lower bounds for the Gallager insertion channel in Fig.~\ref{fig:bottom_plot}. For the first upper bound (UB), we consider a genie-aided system in which the insertion positions are revealed to the receiver as side information. The resulting channel model is equivalent to a binary erasure channel with erasure probability $\alpha$, whose capacity is $1-\alpha$. This can be considered an upper bound for the Gallager insertion channel due to the use of side information. For the second UB, we adopt the approach proposed in \cite{fertonani2010bounds} for the channels with insertions, deletions and substitutions and apply it to the Gallager insertion channel. This approach is also genie-aided, but only partial side information is provided to the receiver by dividing the input sequence into length-$L$ subsequences and revealing the length of the corresponding output subsequences, i.e., the receiver knows exactly how many bits are inserted in each subsequence. Clearly, this subchannel is memoryless and has finite input/output alphabets, and hence its capacity can be numerically evaluated using the Blahut-Arimoto algorithm (BAA) by providing the probability of each input/output pair. In this way, an upper bound for the channel capacity is obtained. In Fig.~\ref{fig:bottom_plot}, the corresponding UB is shown for $L=8$. Since the revealed side information is more limited in this case, the resulting UB is, as expected, tighter compared to the first one.

For one of the lower bounds, Gallager~\cite{gallager_key} considered the use of convolutional codes over channels with synchronization errors and derived an expression that represents an achievable rate (or equivalently, a capacity lower bound) for channels with insertion, deletion, and substitution errors. By equating the probabilities of error types other than insertion error to zero, we also provide this result as a lower bound in Fig.~\ref{fig:bottom_plot}. For the second lower bound, we consider the bound given in~\cite{6516943}, which provides an analytical lower bound on the capacity of the binary Gallager insertion channel by considering independent uniformly distributed inputs and computing lower bounds on the mutual information between the input and output sequences. This bound outperforms the one in~\cite{gallager_key} for a wide range of insertion probabilities. We note that our provided approximation fits well between the given lower and upper bounds for the Gallager insertion channel model, as expected.

For the simple insertion channel, we compute the upper bound using the same genie-aided partial side-information approach as for the Gallager insertion channel, combined with the Blahut–Arimoto algorithm. The result for $L=8$ is shown in Fig.~\ref{fig:bottom_plot2}.

The paper is organized as follows. Section \ref{sec:prelim} introduces the preliminaries and the channel models. The maximum rate for the described insertion channel models is decomposed into entropy terms in Section \ref{sec:IXnYn}. The capacity approximation for the simple insertion channel and the Gallager insertion channel are provided in Sections \ref{sec:firstCh} and \ref{sec:GalCh}, respectively. The paper is concluded in Section \ref{sec:disc}.

\textit{Notation:} Throughout the paper, we use base-two logarithms. Additionally, we note that as $\alpha$ approaches 0, we have
\begin{equation*}
\log (1-\alpha) \approx -\alpha \log(e),
\end{equation*}
where $e$ represents Euler's number.

\section{Preliminaries} \label{sec:prelim}
We consider an insertion channel \( W_n \) which takes \( n \)-bit binary vector \( X^n \) as an input and outputs \( Y(X^n) \). We note that the inserted bits are independent of both other insertions and the original input bits. Specifically, we focus on the following two specific channel models in this paper: \vspace{0.1cm}

\noindent \textbf{1) Simple Insertion Channel:} There is a random bit insertion after each bit with probability \( \alpha \). We define the insertion pattern for this channel model by the vector \( A^n \in \{0,1\}^n \), whose \( i \)-th element is defined by
\begin{align*}
    A_i= 
    \begin{cases}
        1,& \text{if there is an insertion after the }i\text{-th bit of }X^n\\
        0,              & \text{otherwise}
    \end{cases}
\end{align*}
for \( i \in \{1,2, \cdots, n\} \).
The corresponding inserted bit vector is denoted by \( B^n  \in \{0,1\}^n \) whose \( i \)-th element is 
\begin{align*}
    B_i= 
    \begin{cases}
        0,& \text{if } A_i = 0\\
        U_i,              & \text{if } A_i = 1
    \end{cases}
\end{align*}
where \( U_i \) are i.i.d. \(\text{Ber}(1/2) \) random variables. Note that value of $B_i$ for $A_i = 0$ is not important and it is arbitrarily set to 0. 

\vspace{0.2cm}

\noindent \textbf{2) Gallager Insertion Channel:} Each bit is replaced by two independent random bits with probability \( \alpha \), as described in \cite{gallager_key}. The insertion pattern for this model is defined by  \( A^n \in \{0,1\}^n \) whose \( i \)-th element is  
\begin{align*}
A_i=
\begin{cases}
1, & \text{if the }i\text{-th bit of }X^n\text{ is replaced } \\
& \text{with two random bits}\\
0, & \text{otherwise}
\end{cases}
\end{align*}
for \( i \in \{1,2, \cdots, n\} \). For this channel model, we define two insertion bit vectors \( B^n = (B^{1,n},B^{2,n}) \) where \( B^{j,n} \in \{0,1\}^n \) with \( j \in \{1,2\} \) and their \( i \)-th element is defined similarly to that of the insertion channel model, i.e., 
\begin{align*}
    B^j_i= 
    \begin{cases}
        0,& \text{if } A_i = 0\\
        U_{i,j},              & \text{if } A_i = 1
    \end{cases}
\end{align*}
for \( j \in \{1,2\} \), where \( U_{i,j} \) are i.i.d. and \(\sim \text{Ber}(1/2) \).

Note that for both channel models, the expected length of the output vector \( Y(X^n) \) is \( \mathbb{E}[n'] = n + \alpha n \), and we will use the notation \( Y^{n'} \) for the output vector.

In \cite{dobrushin1967shannon}, Dobrushin showed that the maximum rate for reliable communication can be formulated by the maximal mutual information per bit for memoryless channels with synchronization errors. It is also shown that information stability holds and Shannon capacity exists. This approach can be used to express the capacity of the insertion channels above, as stated in the following theorem.
\begin{theorem}  \label{kanoria_theorem} 
(Taken from \cite{dobrushin1967shannon}) For the memoryless channels with finite expectation of output vector length, consider the definition
    \begin{equation}
    C_{n} \equiv \frac{1}{n} \max_{p_{X^n}} I(X^n; Y(X^n)).
\end{equation}
Then, the following limit exists and defines the capacity of the channels with synchronization errors
\begin{equation}
    C \equiv \lim_{n \rightarrow \infty} C_n = \inf_{n \geq 1} C_n.
\end{equation}
\end{theorem}

Furthermore, as stated in \cite[Th. 5]{dobrushin1967shannon}, the input $(X_1, X_2, \cdots, X_n)$ can be considered as consecutive coordinates of a stationary ergodic process as in the following lemma.
\begin{lemma} \label{lem_kanoria}
(Taken from \cite{kanoria2013optimal}) With $X_i \in \{0,1\} $ and stationary and ergodic process $\mathbb X = \{X_i\}_{i \in \mathbb Z}$, the limit $I(\mathbb X) = \lim_{n \rightarrow \infty} \frac{1}{n} I(X^n;Y(X^n))$  exists and we have
 \begin{equation}
    C= \sup _{\mathbb X \in {\mathcal S}} I(\mathbb X),
\end{equation}
where $\mathcal{S}$ denotes  the class of stationary and ergodic processes that take binary values.
\end{lemma}
See \cite{kanoria2013optimal} for the proofs of Theorem \ref{kanoria_theorem} and Lemma \ref{lem_kanoria}.

We define the rate achieved by the stationary and ergodic process \( \mathbb{X} \) as:
\begin{equation} \label{rate_IX}
    I(\mathbb{X}) = \lim_{n \rightarrow \infty} \frac{1}{n} I(X^n;Y(X^n)).
\end{equation}

Throughout this paper, we will use the following definition: for the binary process $\mathbb{X}$, the consecutive blocks of 0's and 1's in $X = (X_1, X_2, \cdots, X_n)$ are called runs, and we investigate the capacity of insertion channels with the stationary ergodic input process $\mathbb{X}$, denoted by $\mathcal{S}$. Furthermore, we denote by $L_0$ the length of the run in $\mathbb{X}$ that contains position 0, and by $L_1, L_2, \ldots$ the lengths of the subsequent consecutive runs. {\color{black}{For \(L \in \mathbb{N}\), we define
\begin{equation*}
\mathcal{S}_L = \{\mathbb{X} \in \mathcal{S} : \Pr(L_0 > L) = 0 \},
\end{equation*}
that is, the class of stationary and ergodic binary processes whose run lengths are almost surely
bounded by \(L\).}}

We are interested in studying the capacity of insertion channels for small insertion probabilities. Hence, we focus on an insertion channel with insertion probability 
$\alpha = 0$ (namely, a perfect channel), and note that the capacity of such a channel is 1, and it is achieved with i.i.d. Bernoulli($1/2$) input distribution. We expect that if we consider i.i.d. input distributions for insertion channels with small insertion probabilities, the achievable rate will not deviate from the channel capacity by much. In fact, we will obtain our achievability results by using i.i.d. Bernoulli($1/2$) inputs, and by proving that the rate achieved matches the first two order terms of the actual channel capacity (by coupling the results with relevant converse bounds). 

To prove the converse results, taking a run-length perspective, we will assume that no run exceeds a certain length. We first will show that limiting the run length does not lead to significant loss if the threshold is sufficiently large. Then, we will derive an upper bound on $I(\mathbb{X})$ over all stationary and ergodic processes. The obtained upper bound will match the achievable rates (with i.i.d. inputs) for the two orders terms of the channel capacity, hence proving the channel capacity approximation for the insertion channel model for small insertion probabilities.

\section{Decomposing Rate: Understanding $I(X^n;Y(X^n))$} \label{sec:IXnYn}

In this section, we identify the different terms necessary to determine the maximum rate for both insertion channel models under consideration. First, we define the runs for the input bit vector, which consists of consecutive bits with the same polarity, and a corresponding vector representing the lengths of the output segments (including the inserted bits). Using the channel models, insertion pattern, and inserted bit vector defined in Section \ref{sec:prelim}, we analyze the contributing factors to the capacity of the simple and Gallager insertion channels, which will be computed in detail in subsequent sections.

We use \( X(j) \) to represent the \( j \)-th run in the input vector \( X^n \), while \( Y(j) \) denotes the corresponding sequence of bits in the output vector \( Y^{n'} \) including any inserted bits; i.e., \( Y(j) \) consists of the bits originating from \( X(j) \) together with any inserted bits if an insertion occurs within \( X(j) \).  For instance, consider an input spanning from the $j$-th to the $(j+2)$-th run:
\begin{equation*}
    [0 \ 0 \ 1 \ 1 \ 1 \ 1 \ 0 \ 0 \ 0 ].
\end{equation*}
Clearly, we have
\begin{align*}
    X(j)   &= [0 \ 0], \\
    X(j+1) &= [1 \ 1 \ 1 \ 1], \\
    X(j+2) &= [0 \ 0 \ 0].
\end{align*}
Assume that the insertion channel introduces a `0' after the first and third bits, and a `1' after the eighth bit. With this insertion pattern, the resulting output runs, defined according to our convention, are as follows:
\begin{align*}
    Y(j)   &= [0 \ {\color{red}{0}} \ 0], \\
    Y(j+1) &= [1 \ {\color{red}{0}} \ 1 \ 1 \ 1], \\
    Y(j+2) &= [0 \ 0 \ {\color{red}{1}} \ 0].
\end{align*}
Note that, as in this example, under our output run definition, the output runs may contain bits with different polarities.

We define \( K \) as the vector of \( |Y(j)| \)'s, where \( |\cdot| \) indicates the number of bits in \( Y(j) \). Assuming that there are \( M \) runs in \( X(j) \), one can express this as:
\begin{align}
    X^{n} &= X(1) \dots X(M-1)X(M), \\
     Y^{n'} &= Y(1) \dots Y(M-1)Y(M), \\
    K &= (|Y(1)|, \dots , |Y(M-1)|, |Y(M)|). \label{def_k}
\end{align} 
We have the following expression for the rate given in \eqref{rate_IX} in terms of entropy:
\begin{align} \label{cap1}
    I(X^{n}; Y) & = H(Y) - H(Y | X^{n}) \\
     &\stackrel{(r_1)}{=} H(Y) - H(Y,K | X^{n}) + H(K|X^n,Y),
\end{align}
where \((r_1)\) follows from the chain rule for entropy. 
For the term \( H(Y,K | X^{n}) \), we have
\begin{align} \label{yx}
    H(Y ,K| X^{n}) &\stackrel{(r_2)}{=}  H(Y,K,A^n,B^n |X^n)   - H(A^n,B^n|X^n, Y, K)  \nonumber \\ 
    &\stackrel{(r_3)}{=}  H(Y,K|X^n,A^n,B^n) + H(A^n,B^n|X^n)   - H(A^n,B^n|X^n, Y,K) \nonumber  \\
    &\stackrel{(r_4)}{=}   H(A^n,B^n|X^n) - H(A^n,B^n|X^n, Y,K)  \nonumber \\
    &\stackrel{(r_5)}{=}  H(A^n,B^n) - H(A^n,B^n|X^n, Y,K),  
\end{align}
where \((r_2)\) and \((r_3)\) follow from the chain rule for entropy, \((r_4)\) since \( H(Y,K|X^n,A^n,B^n) = 0 \), and \((r_5)\) due to the independence of \( A^n \) and \( B^n \) with the input \( X^n \).
Consequently, we obtain the expression:
\begin{align} \label{cap2}
    I(X^{n}; Y) =& H(Y) - H(A^n,B^n)  + H(A^n,B^n|X^n, Y,K)+ H(K|X^n,Y).
\end{align}
In this expression, $H(Y)$ quantifies the randomness in the output for all possible input sequences. $H(A^n,B^n)$ represents the uncertainty due to the insertion pattern and inserted bits, while $H(A^n,B^n|X^n, Y,K)$ indicates the ambiguity of the insertion pattern and the inserted bits after observing the input and output sequences, and run lengths. The last term, $H(K|X^n,Y)$, captures the unpredictability of the run lengths after observing the input and output.

\section{Capacity Approximation for the Simple Insertion Channel Model} \label{sec:firstCh}

In this section, we provide a capacity approximation for the first insertion channel model (dubbed the simple insertion channel) by computing the first two terms of the series expansion for small insertion probabilities. Our main result is given as follows.
\begin{theorem}
    Let $C_1(\alpha)$ be the capacity of the simple insertion channel with insertion probability $\alpha$. Then, for small $\alpha$ and any $\epsilon > 0$, we have
    \begin{align}
        C_1(\alpha) = 1 + \alpha \log(\alpha)  +G_1\alpha + \mathcal{O}(\alpha^{3/2-\epsilon})
    \end{align}
 where 
 \begin{align}
     G_1 &=  -\log(e) + \frac{1}{2}\sum_{l=1}^{\infty} 2^{-l-1}l \log l  +  \frac{1}{2}\sum_{a = 1}^\infty  (a+1)2^{-a}h\left(\frac{1}{a+1} \right),
 \end{align}
and it is approximately $0.4901$.
\end{theorem}

To establish this approximation, the first step involves calculating the terms in $I(X^n, Y(X^n))$, as explicitly given in Section \ref{sec:IXnYn} and provided in \eqref{cap2}. We elaborate on these calculations in the subsequent subsections.

\subsection{Calculation of $H(A^n,B^n)$} \label{SecHAB}
We define the process denoting the insertion patterns corresponding to the sequence of insertion channels as $\mathbb{A}$, which is an i.i.d. Bernoulli($\alpha$) process that determines the locations of insertions, independent of the input. With this process and channel model, we represent the corresponding $n$-bit binary vector as $A^n$. Each element of $A^n$, denoted as $A_i$, is set to `1' if there is an insertion after the $i$-th bit of the input vector, and `0' otherwise. For the inserted bits, as described in Section \ref{sec:prelim}, $B^i = 0$ if there is no insertion after the $i$-th bit while $B_i  \sim \text{Ber}(1/2)$ otherwise.

\begin{lemma} \label{lemma:Hab1}
We have  
  \begin{align}  
  \frac{1}{n} H(A^n, B^n)  = h(\alpha) + \alpha,
\end{align}  
where $h(\cdot)$ is the binary entropy function. 
\end{lemma}
\begin{proof}
To compute $H(A^n,B^n)$ in \eqref{cap2}, one can use the independence of the bits:
\begin{equation} \label{hdb}
    H(A^n,B^n) = \sum_{i=1}^n H(A_i, B_i).
\end{equation}
Note that the joint probability mass function (PMF) of $A_i$ and $B_i$ can be expressed as:
\begin{align} \label{probs}
     P(A_i = a, B_i = b)= 
\begin{cases}
    1-\alpha,& \text{if } (a,b) = (0,0)\\
    \frac{\alpha}{2},              & \text{if } (a,b) = (1,0) \\ & \text{or } (a,b) = (1,1) \\
    0,& \text{otherwise}.
\end{cases}
\end{align}
As a result, we have
\begin{align}
    H(A_i, B_i) & = -(1-\alpha)\log(1-\alpha)-\alpha \log(\alpha/2) \nonumber \\
    &= -(1-\alpha)\log(1-\alpha)-\alpha\log(\alpha)+\alpha \nonumber \\
    &= h(\alpha) + \alpha,
\end{align}
where $h(\alpha)$ is the binary entropy function. We can also write
\begin{align}  \label{hdbx}
    H(A^n, B^n)  = nh(\alpha) + n\alpha,
\end{align}
hence the result follows.
\end{proof}

\subsection{Calculation of $H(A^n, B^n | X^n, Y, K)$} \label{sec:Habxyk}

To calculate $H(A^n, B^n | X^n, Y, K)$, we consider a run perspective approach based on the input run structure. First, we introduce a modified insertion process, defined by the modified vectors $\hat{A}^n$ and $\hat{B}^n$. In this modified process, we consider at most a single bit insertion for each run. If there is more than one insertion in a run, all the insertions in that run are reversed (i.e., the run is transmitted without any insertion, and the input and output runs are identical). Note that for small insertion probabilities, typical insertion patterns will have insertions that are sufficiently spaced apart and will contribute to the higher-order terms in the capacity. Multiple insertions within a run will not affect the higher-order terms as will be shown later. Hence, one can obtain an accurate estimation of $H(A^n, B^n | X^n, Y, K)$ when the insertion probability is small (and consequently the capacity).

%

%
We further define stationary processes $ {\mathbb{Z}}^n$ and $ {\mathbb{V}}^n$ to keep track of reversed insertions as follows: 
\begin{equation} \label{eqz}
    {\mathbb{Z}}^n = \hat{\mathbb{A}}^n \oplus  \mathbb{A}^n,
\end{equation}
and 
\begin{equation} \label{eqv}
    {\mathbb{V}}^n = \hat{\mathbb{B}}^n \oplus  \mathbb{B}^n.
\end{equation}
With the modified insertion process, since we leave the runs unchanged if there are no insertions or exactly one insertion in the corresponding run, for these runs, ${\mathbb{Z}}^n$ and ${\mathbb{V}}^n$ consist entirely of 0's. For the remaining runs, the corresponding segments in ${\mathbb{Z}}^n$ and ${\mathbb{V}}^n$ are the same as in ${\mathbb{A}}^n$ and ${\mathbb{B}}^n$, respectively. Therefore, ${\mathbb{Z}}^n$ and ${\mathbb{V}}^n$ represent the segments in ${\mathbb{A}}^n$ and ${\mathbb{B}}^n$ that are reversed due to the modified insertion process, while the segments corresponding to runs with at most one insertion remain unchanged.

As an example, consider the following binary input vector:
\begin{equation*}
    X^n = [0 \ 0 \ \vdots \ 1 \ 1 \ 1 \ 1 \ \vdots \ 0 \ 0 \ 0 ],
\end{equation*}
with insertion pattern and inserted-bit vectors given by
\begin{align*}
    A^n &= [0 \ 0 \ \vdots \ 0 \ 1 \ 0 \ 1 \ \vdots \ 0 \ 1 \ 0 ],\\
    B^n &= [0 \ 0 \ \vdots \ 0 \ 1 \ 0 \ 0 \ \vdots \ 0 \ 1 \ 0 ].
\end{align*}
With this insertion pattern, there are two insertions in the second run of $X^n$, which means that these insertions are reversed with the modified insertion process, i.e., we have
\begin{align*}
    \hat{A}^n &= [0 \ 0 \ \vdots \ 0 \ 0 \ 0 \ 0 \ \vdots \ 0 \ 1 \ 0 ],\\
    \hat{B}^n &= [0 \ 0 \ \vdots \ 0 \ 0 \ 0 \ 0 \ \vdots \ 0 \ 1 \ 0 ],
\end{align*}
and hence the resulting segments of the processes ${\mathbb{Z}}^n$ and ${\mathbb{V}}^n$ are
\begin{align*}
    Z^n &= [0 \ 0 \ \vdots \ 0 \ 1 \ 0 \ 1 \ \vdots \ 0 \ 0 \ 0 ],\\
    V^n &= [0 \ 0 \ \vdots \ 0 \ 1 \ 0 \ 0 \ \vdots \ 0 \ 0 \ 0 ].
\end{align*}
Note that, as already stated in our definition of ${\mathbb{Z}}^n$ and ${\mathbb{V}}^n$, these segments keep track of the insertions that are reversed by the modified insertion process.

In the rest of this section, we first derive a bound on the difference between \(H(\hat{A}^n, \hat{B}^n | X^n, \hat{Y}, \hat{K})\) and \(H({A}^n, {B}^n | X^n, {Y}, {K})\) for the modified insertion process and the original insertion process where $\hat{Y}$ is the output vector when the modified insertion process is applied, and $\hat{K}$ represents the run lengths of the output (defined similarly to \eqref{def_k}, including both the bits in the original input and the inserted bits from the modified insertion process). This represents the approximation error incurred by assuming the modified insertion process instead of the original one. Later, we will show that this bound does not affect the first two-order terms, thereby not impacting our capacity approximation. Next, we compute \(H(\hat{A}^n, \hat{B}^n | X^n, \hat{Y}, \hat{K})\), which represents the ambiguity in the insertion pattern and inserted bits given the input, output, and associated run vector when the modified insertion process is applied. By combining these two steps, we will arrive at an approximation for \(H({A}^n, {B}^n | X^n, {Y}, {K})\).

We begin by examining the difference between \(H(\hat{A}^n, \hat{B}^n | X^n, \hat{Y}, \hat{K})\) and \(H({A}^n, {B}^n | X^n, {Y}, {K})\) for the modified and original insertion processes considering the following pairs:
\begin{align} \label{pair1}
    \left( (X^n, Y, K, A^n, B^n), (X^n, \hat Y, \hat{K}, \hat A^n, \hat B^n) \right),
\end{align}
and
\begin{align} \label{pair2}
    \left( (X^n, Y, K), (X^n, \hat Y, \hat{K} ) \right).
\end{align}
Both of these pairs are in the form of $(T,E)$ such that $T$ is a function of $(E, Z^n, V^n)$ and $E$ is a function of $(T, Z^n, V^n)$. 
Note that even though there is no exact formulation for $T$ and $E$, for the pair~\eqref{pair1}, we have $T = (X^n, Y, K, A^n, B^n)$ and $E = (X^n, \hat Y, \hat{K}, \hat A^n, \hat B^n)$, and $T$ is a function of $(E, Z^n, V^n) = (X^n, \hat Y, \hat{K}, \hat A^n, \hat B^n, Z^n, V^n)$; that is, $T$ can be recovered from $(X^n, \hat Y, \hat{K}, \hat A^n, \hat B^n)$ once we also have the information of $Z^n$ and $V^n$, which describe the reversed insertions as explained previously. One can extend this analogy in the other direction (i.e., vice versa), as well as to the second pair given in~\eqref{pair2}.

From the chain rule of entropy, we obtain:
\begin{align}
    H(T) &= H(T, E) - H(E | T), \\
    H(E) &= H(T, E) - H(T | E).
\end{align}
Taking the difference, we get
\begin{equation}
    H(T) - H(E) = H(T | E) - H(E | T).
\end{equation}
By the chain rule, we can also write
\begin{align}
    H(T | E) = H(T|E,Z^n, V^n)+&H(Z^n, V^n|E) - H(Z^n, V^n|E,T),
\end{align}
Since $ T $ can be written as a function of $ (E, Z^n, V^n) $, we have $H(T|E,Z^n, V^n) = 0$, and 
\begin{align}
    H(T | E) \leq H(Z^n, V^n).
\end{align}
Similarly, one can follow the same steps for $H(E | T)$ resulting in the same upper bound.
Thus, we have:
\begin{align} \label{hae}
    |H(T) - H(E)| \leq H(Z^n, V^n).
\end{align}
%
%
Therefore, we have:
\begin{align} 
    &|H(X^n, Y, K, A^n, B^n)-H(X^n,  \hat Y,\hat{K}, \hat A^n, \hat B^n)| \leq H(Z^n,V^n) \label{ab1}
\end{align}
For the second pair, we also have:
\begin{equation}  \label{ab2}
    |H(X^n, Y, K)-H(X^n, \hat Y, \hat{K})| \leq H(Z^n,V^n).
\end{equation}
Combining \eqref{ab1} and \eqref{ab2} and using the definition of conditional entropy, we have:
\begin{align} \label{dif_final1}
    |H(A^n, B^n | X^n, Y,K) -H(\hat A^n, \hat B^n &| X^n,  \hat Y, \hat{K})| \leq 2H(Z^n,V^n).
\end{align}
Clearly, the processes $\mathbb{Z}$ and $\mathbb{V}$ are stationary, and we have $2H(Z^n,V^n)/n \leq 2h(z,v)$, where $h(z, v)$ denotes the joint entropy of the random variables $z$ and $v$, whose corresponding probabilities are given below.
Let us calculate the PMF $P(z = a, v = b)$.
\begin{align} \label{pzv}
P(z = 0, v = 0) &= P(z = 0) P(v = 0 | z = 0) = P(z = 0), \nonumber \\
P(z = 0, v = 1) &= P(z = 0) P(v = 1 | z = 0) = 0, \nonumber \\
P(z = 1, v = 0) &= P(z = 1) P(v = 0 | z = 1) = \frac{1}{2} P(z = 1), \nonumber \\
P(z = 1, v = 1) &= P(z = 1) P(v = 1 | z = 1) = \frac{1}{2} P(z = 1). \end{align} 
Without loss of generality, let us focus on the first bit of a run. If there is an insertion at this bit, it will be reversed if at least one more insertion occurs within the run containing this run. For simplicity, we can consider the first run (the run containing the bit at position 0), which has a length of $L_0$.
Conditioned on $L_0$, the probability that $\{z = 1\}$ is given by $\alpha \left(1 - (1 - \alpha)^{L_0 - 1}\right)$. Therefore, we can write 
\begin{align} \label{probZ} 
P(z = 1) = \alpha \left(1 - \mathbb E [(1 - \alpha)^{L_0 - 1}] \right). 
\end{align} 
Thus, the joint PMF $P(z = a, v = b)$ can be written as: 
\begin{align} \label{probZz}
     P(z=a, v=b) = 
\begin{cases}
    & 1- \alpha + \mathbb  E [\alpha(1-\alpha)^{L_0-1}] , \\ 
    & \qquad \qquad \text{if } (a,b) = (0,0) \\
    & \frac{1}{2} \left(\alpha - \mathbb E [\alpha(1-\alpha)^{L_0-1}] \right),             \\
    &  \qquad \text{if } (a,b) = (1,0) \text{ or } (1,1) \\
    & 0 , \qquad \qquad \text{otherwise},  
\end{cases}
\end{align}
We can use this expression to compute $h(z,v)$. 

We further have $(1-\alpha)^{L_0-1} =  1-\alpha(L_0-1) + \mathcal{O}(\alpha^2)$. Hence, the PMF can be rewritten as:
\begin{align} \label{probZ_appr}
     P&(z=a, v=b) \nonumber \\ &= 
\begin{cases}
    1- (\mathbb E [L_0]-1)\alpha^2 + \mathcal{O}(\alpha^3),& \text{if } (a,b) = (0,0) \\
    \frac{1}{2} \mathbb E [L_0-1]\alpha^2 + \mathcal{O}(\alpha^3),              & \text{if } (a,b) = (1,0) \\
    & \text{or } (a,b) = (1,1) \\
    0 ,& \text{otherwise},  
\end{cases}
\end{align}
 which will be used to calculate the dominant terms of $h(z, v)$. This term identifies the loss incurred when restricting ourselves to the modified insertion process, i.e., when we assume there is at most one insertion in a run. Intuitively, these probabilities will not contribute to the dominant terms of the capacity expression. This will be made precise by working out the converse.

As mentioned earlier, to prove the achievability result, we will assume an i.i.d. Bernoulli(1/2) input distribution. Hence, the expectation of $L_0$ will be computed using 
\begin{equation}
    p_{L_0}(l) = 2^{-l},
\end{equation}
which gives $\mathbb{E}[L_0] = 2$. 

For the converse, additional work is required, which will be addressed in the following sections. Initially, we restrict the input run lengths to a specified threshold and show that this constraint does not impact the lower-order terms of the capacity. By deriving an upper bound for the converse and demonstrating its agreement with the achievability results for the first two-order terms, we will finalize the proof for the capacity approximation in the regime of small insertion probabilities.

\begin{lemma} \label{lemma:habxyk1}
       Consider any $\mathbb X \in \mathcal{S}$ such that  $\mathbb{E}[L_0 \log L_0] < \infty$. Then, we have
    \begin{equation}
        \lim_{n \rightarrow \infty } \frac{1}{n}H({A}^n,{B}^n|X^n, {Y}(X^n),K) =  \frac{1}{2}\alpha \mathbb{E} [\log (L_0)] - \eta,
    \end{equation}
    where $ - 2h(z,v) \leq  \eta \leq \frac{1}{2}\alpha^2 \mathbb{E} [L_0 \log (L_0)]+ 2h(z,v)$. Note that $h(z, v)$ can be computed using the probabilities provided in \eqref{pzv}. Also, this quantity will be used explicitly in the achievability and converse results later in the paper.
\end{lemma}

\begin{proof}

The proof follows from \eqref{dif_final1} and Lemma \ref{lemmaABXY_marker} given next.
\end{proof}

\begin{lemma} \label{lemmaABXY_marker}
    Consider any $\mathbb X \in \mathcal{S}$ such that  $\mathbb{E}[L_0 \log L_0] < \infty$. Then, we have
    \begin{equation}
        \lim_{n \rightarrow \infty } \frac{1}{n}H(\hat{A}^n,\hat{B}^n|X^n, \hat{Y}(X^n),\hat{K}) =  \frac{1}{2}\alpha \mathbb{E} [\log (L_0)] - \delta,
    \end{equation}
    where $ 0\leq  \delta \leq \frac{1}{2}\alpha^2 \mathbb{E} [L_0 \log (L_0)]$.
\end{lemma}

\begin{proof}
We follow a similar approach to that of~\cite[Lemma IV.4]{kanoria2010deletion}. The details are given below.

Having information $\hat{K}$ is equivalent to having a marker vector for the last positions of each run. Note that these indices are determined using the positions of the bits in $\hat{Y}(X^n)$ corresponding to the last bit of a run (including the position with an insertion after the last bit if there is such an insertion) in $X$.

The proof consists of the following steps. First, we fix the channel input $x^n$ and any possible output $\hat{y}(x^n)$. Then, we estimate (the $\log$ of) the number of possible realizations of $\hat{A}^n$ and $\hat{B}^n$ that might lead to the input/output pair $(x^n,\hat{y})$ given the marker position vector $\hat k$. The final step is to take the expectation over $(x^n,\hat{y},\hat k)$.

Proceeding from left to right, and using the constraints on both $\hat{\mathbb A}^n$ and $\hat{\mathbb B}^n$, one can map each run of $\hat y$ to a run of $x^n$ given the marker vector $\hat k$. To simplify the exposition, let us focus on the $r$-th run of $x^n$ which consists of `0's. $l_r$ denotes the length of this run.
\begin{equation}
    [{\color{lightgray}{1}} {\color{purple}{\Vert}}  0 {\color{red}{\wedge}} 0 {\color{red}{\wedge}} 0 {\color{red}{\wedge}} 0 {\color{red}{\wedge}} {\color{purple}{\Vert}} {\color{lightgray}{1}} ]_{1 \times (l_r+2)},
\end{equation}
${\color{red}{\wedge}}$ denotes the possible positions for insertions inside the run, and ${\color{purple}{\Vert}}$ denotes the marker positions for the edge of the run. Clearly, there are $l_r$ possible positions for insertions. Note that, since we use $\hat{\mathbb A}^n$ and $\hat{\mathbb B}^n$, there is at most a single insertion over each run.

For this example, there are three cases:
\begin{enumerate}
\item \textbf{No insertion: } Since there is no insertion, there is a one-to-one mapping between the run of $x^n$ and $\hat{y}$. Hence, there is no ambiguity and no contribution to $H(\hat A^n,\hat B^n|x^n, \hat y,\hat k)$.
    \item \textbf{Insert bit `0':} There are $l_r$ possible locations (each of them being equally likely) for insertion in $\hat A$ given the marker vector $\hat K$. 
        
    There is no additional ambiguity due to $\hat B$ (the corresponding bits of $\hat B$ will be all zero regardless of $l_r$ possible insertion positions). 

    Hence, there will be a contribution of $\log(l_{r})$ to $H(\hat A^n,\hat B^n|x^n, \hat y, \hat k)$.


        \item \textbf{Insert bit `1':} Since the original run of $x^n$ consists of `0's, inserting `1' will reveal all the information regarding $\hat A$ and $\hat B$ given the marker vector $k$. Hence, there is no contribution to $H(\hat A^n,\hat B^n|x^n, \hat y , \hat k)$.
\end{enumerate}

Note that, a similar argument can be made for other runs of `0's or `1's.


\noindent The total contributions due to case 2 previously explained (insertion of `0' into runs of 0's, and insertion of `1' into runs of 1's) can be expressed as a summation over all bits in the input for the stationary and ergodic process and given by:
\begin{align} \label{abxy2_marker}
     \frac{1}{2} \sum_{i = 1}^n \hat A_i \log(l_{(i)}), 
\end{align}
where $l_{(i)}$ is the length of the run containing the $i$-th bit of $x^n$ and $\hat A_i$ is the $i$-th bit of the modified insertion pattern. Note that if there is no insertion for a certain bit, $\hat A_i = 0$ and there will be no contribution to the summation.

We further have
\begin{align}
    P(\hat A_i = 1) &= P\left(\text{Having one insertion after the }i\text{-th  bit }\right) \nonumber \\
    &= P\left( \hat A = [ {\color{lightgray}{0}} \vdots \cdots 0's \cdots  1 \cdots 0's \cdots  \vdots {\color{lightgray}{0}} ]_{1 \times (l_r+2)}\right) \nonumber \\
            &= (1-\alpha)^{l_r-1}\alpha.
    \end{align}
Hence, 
\begin{equation} \label{pai}
    P(\hat A_i = 1) = (1-\alpha)^{l_{(i)}-1}\alpha ,
\end{equation}
and we can write $P(\hat A_i = 1)  \in (\alpha-(l_{(i)}-1)\alpha^2,\alpha)$, which can be rewritten as
\begin{equation} \label{pai}
    P(\hat A_i = 1)  \in (\alpha-l_{(i)}\alpha^2,\alpha).
\end{equation}

Taking the expectation of \eqref{abxy2_marker}, we obtain
\begin{align} 
    \frac{1}{2} \sum_{i = 1}^n \left(1 \cdot P(\hat A_i = 1) + 0 \cdot P(\hat A_i = 0) \right) \mathbb{E}[\log(l_{(i)})] 
    = \frac{1}{2} \sum_{i = 1}^n P(\hat A_i = 1) \mathbb{E}[\log(l_{(i)})].
\end{align}
Letting $n \rightarrow \infty$ and using the bound given in \eqref{pai}, we can equivalently write
\begin{equation} \label{hatdbxy_marker2}
\lim_{n \rightarrow \infty} \frac{1}{n} H(\hat{A}^n,\hat{B}^n|X^n, \hat{Y}, \hat{K}) 
= \frac{1}{2} \alpha \mathbb{E}[\log(L_0)] - \delta,
\end{equation}
where $0 \leq \delta \leq \frac{1}{2}\alpha^2 \mathbb{E}[L_0 \log(L_0)]$, which completes the proof of Lemma~\ref{lemmaABXY_marker}.

\end{proof}

\subsection{Calculation of $H(K|X^n,Y)$} \label{SecHKXY}

\begin{lemma} \label{lemma:hkxy_first}
For a simple insertion channel with {\color{black}{any $\epsilon > 0$, 
there exists $\alpha_0 \equiv \alpha_0(\epsilon) > 0$ such that for any $\alpha < \alpha_0$
and any stationary and ergodic input process $\mathbb{X} \in \mathcal{S}_{\lfloor 1/\alpha \rfloor}$}} satisfying $H(\mathbb{X}) > 1 - \alpha^{\gamma}$, where $\gamma > \frac{1}{2}$, we have:
    \begin{align} 
 \lim_{n \rightarrow \infty} \frac{1}{n}  H(K|X^n,Y) =  \frac{\alpha}{2} \sum_{a = 1}^\infty   (a+1) 2^{-a}   h\left(\frac{1}{a+1} \right) + \epsilon_1,
 \end{align}
 where 
 \begin{equation}
     - \alpha^{1+\gamma/2-\epsilon}- 4\alpha^{1+\gamma-\epsilon/2} - \frac{\alpha^2}{2}\sum_{r_j = 1}^\infty \sum_{r_{j+1} = 1}^\infty  (r_j + r_{j+1}) (r_{j+1}+1)h\left(\frac{1}{r_{j+1}+1} \right) \leq \epsilon_1 \leq  4\alpha^{1+\gamma-\epsilon/2} + \alpha^{1+\gamma/2-\epsilon},
 \end{equation}
 with $\epsilon> 0$. 
\end{lemma}

\begin{proof}

To compute $H(K|X^n,Y)$, we begin by defining a perturbed insertion process, denoted as $\check{\mathbb{A}}$. For the input process $\mathbb{X}$, we introduce $\check{\mathbb{Z}}$, whose elements correspond to two consecutive input runs with at most one insertion, equated to `0'. Otherwise, the elements mirror those of the original insertion process ${\mathbb{A}}$. To clarify further, for two consecutive input runs $\{S_i, S_{i+1}\}$, we define $\check{\mathbb{Z}}^i$ as a binary process with all zeros when $\{S_i, S_{i+1}\}$ has at most one insertion. Otherwise, $\check{Z}_l^i = 1$ if $X_l \in S_i$ and $A_l = 1$, $\forall i$.
Utilizing the definition of $\check{Z}_l^i$, we further define $\check{\mathbb{Z}}$ in the following way: 
\begin{align}
     \check{Z}_l= 
\begin{cases}
    1,& \text{if } \exists i \text{ such that } \check{Z}_l^i = 1\\
    0,              & \text{otherwise.}
\end{cases}
\end{align}
This perturbed process yields a new insertion pattern defined as:
\begin{equation}
    \check{\mathbb{A}} \equiv \mathbb{A} \oplus \check{\mathbb{Z}}.
\end{equation}
A sample of the runs between $i$ and $i+4$ is provided in Table \ref{tab:samplerun}, showing the changes in the perturbed insertion process compared to the original insertion process. 

\begin{table}[]
\centering \caption{An example illustrating the input runs alongside the original and perturbed insertion processes.}\label{tab:samplerun}
\begin{tabular}{|c|c|c|c|c|c|}
\hline
run index:  & $i$ & $i+1$ & $i+2$ & $i+3$ & $i+4$ \\ \hline
$S$         & 000 & 1111  & 00    & 111   & 0000  \\ \hline
$A$         & 010 & 1000  & 00    & 001   & 0000  \\ \hline
$\check{Z}$ & 010 & 0000  & 00    & 000   & 0000  \\ \hline
$\check{A}$ & 000 & 1000  & 00    & 001   & 0000  \\ \hline
\end{tabular}
\end{table}

Regarding the term $H(K|X^n,Y)$, we first calculate the entropy using the perturbed process, followed by bounding the difference between the original insertion process and the perturbed process. 

%
The output of this perturbed process is denoted by $\check{Y}(X^n)$, and we define $\check{K}$ as follows:
\begin{equation}
      \check{K} = (|\check{Y}(1)|, \dots , |\check{Y}(M-1)|, |\check{Y}(M)|),
\end{equation}
where $\check{Y}(j)$ denotes the bits in the output corresponding to the $j$-th run of the input (including the original bits and insertions due to the perturbed process). Applying the chain rule, we express $H(\check{K}|X^n,\check{Y})$ as:
\begin{align} \label{hkxy1}
H(\check{K}|X^n,\check{Y}) & = H\left( |\check{Y}(1)|, \dots , |\check{Y}(M-1)|, |\check{Y}(M)| \Big| X^n, \check{Y} \right) \nonumber \\
   & = \sum_{j=1}^M H\bigg( |\check{Y}(j)| \Big|  |\check{Y}(1)|, \dots , |\check{Y}(j-1)|, \nonumber \\
   & \qquad  X(1)X(2) \cdots X(M), \check{Y}(1)\check{Y}(2) \cdots \check{Y}(M) \bigg) ,
\end{align}
where \( X(1)X(2) \cdots X(M) \) and \( \check{Y}(1)\check{Y}(2) \cdots \check{Y}(M) \) represent concatenations of \( X(j) \)'s and \( \check{Y}(j) \)'s, respectively. Therefore, \( X^n = X(1)X(2) \cdots X(M) \) and \( \check{Y} = \check{Y}(1)\check{Y}(2) \cdots \check{Y}(M) \).

In \eqref{hkxy1}, the condition on \( |\check{Y}(1)|, \dots , |\check{Y}(j-1)| \), \( X(1)X(2) \cdots X(j-1) \), and \( \check{Y}(1)\check{Y}(2) \cdots \check{Y}(j-1) \) can be omitted. Hence, equivalently, we obtain:
\begin{align} \label{hkxy1x}
H(\check{K}|X^n,\check{Y}) = \sum_{j=1}^M H\bigg( |\check{Y}(j)| & \Big|  X(j) \cdots X(M),  \check{Y}(j) \cdots \check{Y}(M) \bigg).
\end{align}

The proof is completed by computing $H(\check{K}|X^n,\check{Y})$ in \eqref{hkxy1x} and establishing a bound for $|H({K}|X^n,{Y})- H(\check{K}|X^n,\check{Y})|$, as given in the Lemmas \ref{lem:checkK} and \ref{lem_Kdif}, respectively.
    
\end{proof}

\begin{lemma} \label{lem:checkK}
For a simple insertion channel using the perturbed insertion process with any $\epsilon > 0$, 
{\color{black}{there exists $\alpha_0 \equiv \alpha_0(\epsilon) > 0$ such that for any $\alpha < \alpha_0$}}
and any stationary and ergodic input process $\mathbb{X} \in \mathcal{S}_{\lfloor 1/\alpha \rfloor}$ 
satisfying $H(\mathbb{X}) > 1 - \alpha^{\gamma}$, where $\gamma > \frac{1}{2}$, we have:
    \begin{align} 
\lim_{n \rightarrow \infty} \frac{1}{n}  H(\check K|X^n, \check Y) =  \frac{\alpha}{2} \sum_{r_{j+1} = 1}^\infty   (r_{j+1}+1) 2^{-r_{j+1}}   h\left(\frac{1}{r_{j+1}+1} \right) - \epsilon_2 + \epsilon_3,
\end{align}
which is approximately $\approx 1.2885\alpha - \epsilon_2 + \epsilon_3$ where
\begin{align} \label{eps}
   0 \leq \epsilon_2 \leq  \frac{\alpha^2}{2}\sum_{r_j = 1}^\infty \sum_{r_{j+1} = 1}^\infty & (r_j + r_{j+1}) (r_{j+1}+1) h\left(\frac{1}{r_{j+1}+1} \right),
\end{align}
and
\begin{equation}
   - \alpha^{1+\gamma/2-\epsilon} \leq \epsilon_3 \leq \alpha^{1+\gamma/2-\epsilon}.
\end{equation}
\end{lemma}
\begin{proof}
We define:
\begin{equation}
    t_j \equiv H\left( |\check{Y}(j)| \Big|  X(j) \cdots X(M), \check{Y}(j) \cdots \check{Y}(M) \right)
\end{equation}

With a slight change in notation, let us define $\bar{Y}(j')$ as the runs in $\check Y$, meaning that $\bar{Y}(j')$ is directly determined by the bits of $\check Y$. Consequently, we can rewrite $t_j$ as:
\begin{equation}
    t_j \equiv H\left( |\check Y(j)| \Big|  X(j) \cdots X(M), \bar{Y}(j') \cdots \bar{Y}(M') \right).
\end{equation}

We note that our procedure proceeds from left to right. Since insertions may introduce new runs under our definition of $\bar{Y}$, the indices of $\check{Y}(j)$ and $\bar{Y}(j')$ may not coincide, even though the corresponding runs themselves do match. However, by starting from the left, one can cancel the runs that are resolved on that side and then continue the process accordingly. For example, the first bit of $\check{Y}(j)$ will correspond to the first bit of $\bar{Y}(j')$; hence, although there may be an index mismatch between $j$ and $j'$, the bits themselves are aligned and represent the same run segment.


Consider four cases:


1. When $|\bar{Y}(j')| = |X(j)|$ and $|\bar{Y}(j'+1)| = |X(j+1)|$: In this scenario, there is no insertion in $X(j)$. Thus, $t_j = 0$, since we have sufficient information to determine $|\check Y(j)|$, given $X(j) \cdots X(M)$.

2. When $|\bar{Y}(j')| < |X(j)|$: Here, we observe a single-bit insertion inside $X(j)$ with the inserted bit having the opposite polarity, for example, inserting a `0' bit into a run of `1's.
    
    We can directly determine $|\check Y(j)|$ using $\bar{Y}(j')$, $\bar{Y}(j'+1)$, and $\bar{Y}(j'+2)$. Hence, $t_j = 0$.
    
    Note that $|\bar{Y}(j'+1)| = 1$, and the corresponding bit has the opposite polarity to $\bar{Y}(j')$ and $\bar{Y}(j'+2)$.

3. When $|\bar{Y}(j')| > |X(j)|$ and $|\bar{Y}(j'+1)| = |X(j+1)|$: Here, we observe a single-bit insertion inside $X(j)$ with the inserted bit having the same polarity, for example, inserting a `0' bit into a run of `0's.
    
   Similar to the previous case,  we can directly determine $|\check Y(j)|$ and $t_j = 0$.

4. When $|\bar{Y}(j')| = |X(j)|$ and $|\bar{Y}(j'+1)| = |X(j+1)|+1$: In this case, a single insertion occurs in $X(j)X(j+1)$, and we cannot distinguish between two different scenarios. Consider the following example:
    \begin{align}
        X(j)X(j+1) & = [1 \ 1\ 1\ 0\ 0\ 0\ 0] \\
        \bar{Y}(j')\bar{Y}(j'+1) & = [1 \ 1\ 1\ 0\ 0\ 0\ 0\ 0]
    \end{align}
    Here, we cannot determine whether the `0' bit is inserted at the end of $X(j)$ (the last possible position for insertion in $X(j)$) or at any of the possible $|X(j+1)|$ locations in $X(j+1)$. This scenario can be covered by the following two cases:
    \begin{itemize}
        \item Case $\mathcal{V}_1$: If the `0' bit is inserted at the end of $X(j)$: 
        \begin{align}
            |\check Y(j)| &= |\bar{Y}(j')| + 1 = |X(j)| + 1\\
            |\check Y(j+1)| &= |\bar{Y}(j'+1)| - 1 = |X(j+1)| 
        \end{align}

        This case has a probability:
    \begin{equation}
        P(\mathcal{V}_1) = \frac{1}{2} \alpha (1- \alpha)^{|X(j)|+|X(j+1)|-1}.
    \end{equation}

        \item Case $\mathcal{V}_2$: If the `0' bit is inserted at any of the possible $|X(j+1)|$ locations in $X(j+1)$,
                \begin{align}
            |\check Y(j)| &= |\bar{Y}(j')|  = |X(j)|   \\
            |\check Y(j+1)| &= |\bar{Y}(j'+1)| +1 = |X(j+1)| +1 
        \end{align}

                This case has a probability:
    \begin{equation}
        P(\mathcal{V}_2) = \frac{1}{2} \alpha(1- \alpha)^{|X(j)|+|X(j+1)|-1} |X(j+1)|.
    \end{equation}

    \end{itemize}
    Combining both cases for $|\check{Y}(j)| = |X(j)|$ and $|\check{Y}(j)| = |X(j)|+1$, and defining $r_j \equiv |X(j)|$, we can write:
    \begin{align} \label{yj}
            |\check{Y}(j)| = 
\begin{cases}
    r_j  + 1, &   \text{with probability }   \frac{1}{2}\alpha(1-\alpha)^{r_j+r_{j+1}-1} \\
    r_j,              & \text{with probability }     \frac{1}{2}\alpha(1-\alpha)^{r_j+r_{j+1}-1}  r_{j+1}
\end{cases}
    \end{align}

Note that the only ambiguity in $|\check {Y}(j)|$ arises from \eqref{yj}, as all the previously described scenarios provide the necessary information for correct estimation. Hence, conditioned on case 4, we can normalize the probabilities as follows:
    \begin{align} \label{yj2}
            |\check{Y}(j)| = 
\begin{cases}
   r_j + 1,& \text{with probability } \frac{1}{r_{j+1}+1} \\
    r_j,              & \text{with probability } \frac{r_{j+1}}{r_{j+1}+1} 
\end{cases}
    \end{align}

Consequently, we have:
\begin{equation}
    t_j = h\left(\frac{1}{r_{j+1}+1} \right).
\end{equation}
Thus, the expected contribution of this term to the sum is:
\begin{align} \label{sum_hk}
    \sum_{r_j = 1}^\infty \sum_{r_{j+1} = 1}^\infty & \frac{1}{2}\alpha(1-\alpha)^{r_j+r_{j+1}-1} (r_{j+1}+1)p_{L(2)}(r_j,r_{j+1})h\left(\frac{1}{r_{j+1}+1} \right),
\end{align}
where $p_{L(k)}(l_1, \dots, l_k)$ is the joint distribution of input run lengths for $k$ runs.

It is worth noting that\footnote{We have $(1-\alpha)^x \geq 1-\alpha x$ for $\alpha \rightarrow 0$ and $x \geq 1$.} $(1-\alpha)^{r_j+r_{j+1}-1} \in \left(1-\alpha(r_j+r_{j+1}), 1\right)$, hence \eqref{sum_hk} can be rewritten as:
\begin{align} \label{plk_1}
 \frac{\alpha}{2}\sum_{r_j = 1}^\infty \sum_{r_{j+1} = 1}^\infty  & (r_{j+1}+1) p_{L(2)}(r_j,r_{j+1})  h\left(\frac{1}{r_{j+1}+1} \right) - \epsilon_2,
\end{align}
with
\begin{align} \label{epsilon2}
   0 \leq \epsilon_2 \leq  \frac{\alpha^2}{2}\sum_{r_j = 1}^\infty \sum_{r_{j+1} = 1}^\infty & (r_j + r_{j+1}) (r_{j+1}+1)  h\left(\frac{1}{r_{j+1}+1} \right),
\end{align}
which is obtained by using the upper bound $p_{L(2)}(r_j,r_{j+1}) \leq 1$ in \eqref{epsilon2}.

Note that in \eqref{plk_1}, the optimal distribution of the input run lengths $p_{L(2)}$ is used. To establish a bound for this term, as in \cite[Lemma V.5]{kanoria2013optimal}, let us assume \( p^*_{L(k)}(l_1, \dots, l_k) = 2^{- \sum_{i=1}^k l_i} \). Consider any positive integer \( k \), \( \kappa < \infty \), and \( H(\mathbb{X}) > 1 - \alpha^\gamma \) with \( \gamma > 1/2 \). 
From \cite[Lemma V.5]{kanoria2013optimal}, we have
\begin{equation}
    \sum_{l_1=1}^\infty \cdots \sum_{l_k=1}^\infty \left| p_{L(k)}(l_1, \dots, l_k) - p^*_{L(k)}(l_1, \dots, l_k) \right| \leq \kappa \sqrt{k} \alpha^{\gamma/2},
\end{equation}
which implies that each term inside the summation also satisfies the following inequality, as all terms are nonnegative:
\begin{align} \label{plk_2}
    \left| p_{L(k)}(l_1, \dots, l_k) - p^*_{L(k)}(l_1, \dots, l_k) \right| \leq \kappa \sqrt{k} \alpha^{\gamma/2}.
\end{align}
With  $\mathbb{X} \in \mathcal{S}_{\lfloor 1/\alpha \rfloor}$, for the summation term in \eqref{plk_1}, we define $l \equiv \lfloor 4 \log (1/\alpha) \rfloor)$. 
{\color{black}{Note that we choose $\ell = \lfloor 4\log(1/\alpha)\rfloor$ because, under the
constraint $H(X) > 1 - \alpha^{\gamma}$ with $\gamma > 1/2$, the
run-length distribution has an exponential tail as stated in \cite[Lemma V.3]{kanoria2013optimal}. Consequently, terms involving runs longer than $\ell$ contribute only to higher-order terms and are therefore negligible in the
overall asymptotic expansion.
}}

We have
\begin{align} \label{plk_1x}
 \frac{\alpha}{2}\sum_{r_j = 1}^l & \sum_{r_{j+1} = 1}^l   (r_{j+1}+1) p_{L(2)}(r_j,r_{j+1})  h\left(\frac{1}{r_{j+1}+1} \right) \nonumber \\
=& \frac{\alpha}{2}\sum_{r_j = 1}^{\infty} \sum_{r_{j+1} = 1}^{\infty}   (r_{j+1}+1) 2^{-r_j -r_{j+1}}  h\left(\frac{1}{r_{j+1}+1} \right) - \epsilon_3,
\end{align}
resulting in
\begin{align} \label{hkxy_last}
 \lim_{n \rightarrow \infty} \frac{1}{n}  H(\check K|X^n,Y) =  A_1 \alpha - \epsilon_2 + \epsilon_3,
\end{align}
where
\begin{align}
    A_1 &= 
    \frac{1}{2}\sum_{r_j = 1}^\infty \sum_{r_{j+1} = 1}^\infty  (r_{j+1}+1) 2^{-r_j} 2^{-r_{j+1}} h\left( \frac{1}{r_{j+1}+1} \right) \nonumber \\
    &= 
    \frac{1}{2}\sum_{r_{j+1} = 1}^\infty (r_{j+1}+1)2^{-r_{j+1}} h\left( \frac{1}{r_{j+1}+1} \right) ,
\end{align}
\begin{equation}
   - \alpha^{1+\gamma/2-\epsilon} \leq \epsilon_3 \leq \alpha^{1+\gamma/2-\epsilon},
\end{equation}
and $\epsilon_2$'s range is defined in \eqref{epsilon2}. 

Note that we have $A_1 \approx 1.2885$.
\end{proof}

\begin{lemma} \label{lem_Kdif}
For a simple insertion channel model with a stationary and ergodic input process $\mathbb{X} \in \mathcal{S}_{\lfloor 1/\alpha \rfloor}$ satisfying $H(\mathbb{X}) > 1 - \alpha^{\gamma}$, where $\gamma > \frac{1}{2}$, and for any $\epsilon > 0$, the difference between $H(K|X^n,Y)$ and $H(\check{K}|X^n, \check{Y})$, corresponding to the original insertion and perturbed insertion processes, respectively, satisfies:
\begin{align}
\lim_{n \rightarrow \infty} \frac{1}{n}|H(K|X^n,Y) - H(\check{K}|X^n, \check{Y})| \leq 4 \alpha^{1+\gamma-\epsilon/2}.
\end{align}
\end{lemma}
\begin{proof}
The proof strategy mirrors that of \cite[Lemma V.18]{kanoria2013optimal}. We utilize the perturbed insertion process $\check{Z}$ and define $U(X^n, A^n, Z^n) \in \{t, 0, 1\}^{|{Y}|}$ for each bit in the output of size $|Y|$. Sequentially, for each bit in $Y$, we define the elements of $U$ based on the corresponding bit in $\check{Y}$ and the perturbed insertion process as follows:
\begin{itemize}
    \item For each bit present in both $Y$ and $\check{Y}$, the corresponding element of $U$ is defined as $t$. This case represents the bits originally existing in input $X$ and the insertions with the perturbed insertion process.
    \item For each bit present in $Y$ but not in $\check{Y}$, we assign the corresponding bit in $U$ as either `0' or `1' based on the corresponding bit in $Y$. This case represents the reversed insertions due to the difference between the original insertion process and the perturbed insertion process.
\end{itemize}

Clearly, we have 
\begin{align}
    (X^{n}, Y) & \mathrel{\mathop {\longleftrightarrow}\limits ^{U}} (X^{n}, \check {Y}) \nonumber \\
    (X^{n}, Y, K) & \mathrel{\mathop {\leftarrow \joinrel\relbar\joinrel\relbar\joinrel\rightarrow}\limits ^{(U, Z)}} (X^{n}, \check {Y}, \check {K}).
\end{align}
It is evident that we have:
\begin{align}
|H(\check{K}(X^{n}) | X^{n}, \check{Y}(X^{n})) - H&({K}(X^{n}) | X^{n}, {Y}(X^{n})) |  \leq 2H(U) + H(Z).
\end{align}

With the perturbed process, we further define $\check{z} = P(\check{Z}_j = 1)$ for each $j$, which is the probability of having a reversed insertion for the $j$-th index. The number of insertions reversed in a random run is at most 
\begin{equation}
    \alpha^2 \sum_{l_0, l_1} p_{{L(2)}}(l_0, l_1)(l_0 + l_1 )^2, 
\end{equation}
in expectation, where $p_{(2)}(l_0, l_2)$ is the joint distribution of input run lengths for $2$ runs. This expression can be bounded above by $4\alpha^2 \mathbb{E}[{L}^2]$.

As in \cite[Lemma V.3]{kanoria2013optimal}, there exists $\alpha_0$ such that the following occurs: Consider any $\gamma > 1/2$, and define $\ell \equiv \lfloor 2 \gamma \log(1/\alpha) \rfloor$. For all $\alpha < \alpha_0$, if $\mathbb{X} \in \mathcal{S}$ is such that $H(\mathbb{X}) > 1 - \alpha^{\gamma}$, we have
\begin{equation}
    \sum_{l = \ell}^{\infty} l p_L(l) \leq 20\alpha^\gamma.
\end{equation}
Using \cite[Lemma V.3]{kanoria2013optimal} and ${L} \leq 1/\alpha$ with probability 1, we have 
\begin{equation}
    \mathbb{E}[{L}^2] \leq \frac{1}{\alpha} E[L] \leq 20 \alpha^{\gamma-1},
\end{equation}
which implies the number of reversed insertion in a run can be upper bounded by $80 \alpha^{1+\gamma}$, i.e., we have
\begin{equation} \label{bound1}
    \alpha^2 \sum_{l_0, l_1} p_{{L(2)}}(l_0, l_1)(l_0 + l_1 )^2 \leq 80 \alpha^{1+\gamma}.
\end{equation}
Furthermore, since the length of a run is at least 1, normalizing \eqref{bound1} by 1 yields a bound for $\check{z}$ as well, which is
\begin{equation}
    \check{z} \leq 80 \alpha^{1+\gamma}.
\end{equation}
It follows that 
\begin{equation}
    H(\mathbb{\check{Z}}) \leq h(\check{z}) \leq \alpha^{1+\gamma-\epsilon/2},
\end{equation}
for small enough $\alpha$. 


We further have $u \equiv P(U_j \neq t)$ for all $j$, where $u = \frac{\check{z}}{1+\alpha} \leq \check{z}$. It follows that $H(\mathbb{U}) \leq u + h(u) \leq \alpha^{1+\gamma-\epsilon/2}$ for small enough $\alpha$. Using $H(\check{\mathbb{Z}})$ and $H(\mathbb{U})$, we have 
\begin{align}
\lim_{n \rightarrow \infty} \frac{2H(U) + H(\check{Z})}{n} &= 2(1+\alpha)H(\mathbb{U}) + H(\check{\mathbb{Z}}) \nonumber \\
&\leq 4H(\mathbb{U}) + H(\check{\mathbb{Z}}),
\end{align}
where the inequality is due to $\alpha \in [0, 1]$. Hence, we get
\begin{align}
\lim_{n \rightarrow \infty} \frac{2H(U) + H(\check{Z})}{n} \leq 4\alpha^{1+\gamma-\epsilon/2},
\end{align}
which completes the proof.
\end{proof}

\subsection{Calculations for $H(Y|X)$}
Using the results given in Sections \ref{SecHAB}, \ref{sec:Habxyk} and \ref{SecHKXY}, we obtain the following corollary. 
\begin{corollary} \label{cor1}
For an input process that is stationary and ergodic satisfying $H(\mathbb{X}) > 1 - \alpha^{\gamma}$, where $\gamma > \frac{1}{2}$, we have
\begin{align}
     \lim_{n \rightarrow \infty }  & \frac{1}{n}  H(Y | X^{n}) \nonumber \\
     & = h(\alpha)  + \alpha  \bigg( 1- \frac{1}{2} \mathbb E [\log L_0]  -\frac{1}{2} \sum_{l_j = 1}^\infty \sum_{l_{j+1} = 1}^\infty  (l_{j+1}+1) \nonumber \\
     & \qquad \cdot 2^{-l_j}2^{-l_{j+1}}h\big(\frac{1}{l_{j+1}+1} \big) \bigg) + \zeta,
\end{align}
where $ \zeta = \eta + \epsilon_1$ with
\begin{align}
 & - \alpha^{1+\gamma/2-\epsilon} - 4\alpha^{1+\gamma-\epsilon/2} - \frac{\alpha^2}{2}\sum_{r_j = 1}^\infty \sum_{r_{j+1} = 1}^\infty  (r_j + r_{j+1}) (r_{j+1}+1) h\left(\frac{1}{r_{j+1}+1} \right) \leq \epsilon_1 \nonumber \\
 & \qquad \qquad \qquad \qquad \qquad \qquad \qquad \qquad \qquad \qquad \qquad \qquad \leq  4\alpha^{1+\gamma-\epsilon/2} + \alpha^{1+\gamma/2-\epsilon}, \nonumber \\
    &  -2h(z,v) \leq  \eta \leq \frac{1}{2}\alpha^2 \mathbb{E} [L_0 \log (L_0)]+ 2h(z,v), \nonumber
\end{align}
with $\epsilon> 0$. 
\end{corollary}
\begin{proof}
    We have
    \begin{align}
        H(Y | X^{n}) &=  H(Y,K | X^{n}) - H(K|X^n,Y) \\
        &= H(A^n,B^n) - H(A^n,B^n|X^n, Y,K)  - H(K|X^n,Y).
    \end{align}
    Using Lemmas \ref{lemma:Hab1}, \ref{lemma:habxyk1} and \ref{lemma:hkxy_first} completes the proof. 
\end{proof}

\subsection{{\color{black}{Achievability}}} \label{sec:ach1}
The next lemma proves the achievability of the described insertion channel model, which is obtained by using a Bernoulli i.i.d. input process.

\begin{lemma}[Achievability] \label{lemma:ach}
    Let $\mathbb{X}^*$ be the iid Bernoulli(1/2) process. For any $\epsilon > 0$, we have
     \begin{align}
        C_1(\alpha) = 1 + \alpha \log(\alpha)  +G_1 \alpha +  \mathcal{O}(\alpha^{3/2-\epsilon}).
    \end{align}
\end{lemma}

\begin{proof}
We follow a similar approach to that of~\cite[Lemma III.1]{kanoria2010deletion}.

Since $\mathbb{X}$ is i.i.d. Bernoulli(1/2) with length $n$, it has a run-length distribution $p_L(l) = 2^{-l}$. Additionally, the corresponding output $Y(X^{*,n})$ is also i.i.d. Bernoulli with run length $L_Y = n+\text{Bin}(n,\alpha)$. 

Using the chain rule of entropy, we have
\begin{align}
    H(Y(X^{*,n}),L_Y) &= H(L_Y) + H(Y(X^{*,n})|L_Y) \\
    &= H(L_Y | Y(X^{*,n})) + H(Y(X^{*,n})).
\end{align}
Since $ H(L_Y | Y(X^{*,n})) = 0$, we have 
\begin{equation}
     H(Y(X^{*,n})) = H(L_Y) + H(Y(X^{*,n})|L_Y).
\end{equation}
Note that for $ H(Y(X^{*,n}) \mid L_Y)$, we have
\begin{align}
    H(Y(X^{*,n})|L) &= \sum_{l}H(Y(X^{*,n})|L = l)p_{L_Y}(l) \nonumber \\
    &= \sum_{l} l p_{L_Y}(l)  \label{pl_ref}.
\end{align}
Since $L_Y = n+\text{Bin}(n,\alpha)$, one can equivalently write
\begin{align}
    H(Y(X^{*,n})|L_Y) = n(1+\alpha).
\end{align}
Furthermore, since $L_Y$ has shifted binomial distribution, its entropy is $\mathcal{O}(\log(n))$. Hence, one can conclude that 
\begin{align}
H(Y(X^{*,n})) &= n(1+\alpha) +\mathcal{O}(\log(n)).
\end{align}
Using the estimate of $H(Y|X^{*,n})$ from the Corollary \ref{cor1} with $P(z=a, v=b) = \mathcal{O}(\alpha^2)$ and $\mathbb{E}[L_0 \log L_0 ] < \infty$ completes the proof.
\end{proof}

\subsection{Converse}  \label{sec:conv1}

%
We denote the processes with maximum run length $L$ with probability one by $\mathcal{S}_L$.
Without loss of generality, we restrict ourselves to finite-length runs. 
With the next lemma, we illustrate that restricting ourselves to $\mathcal{S}_{L^*}$ for large enough $L^*$ will not cause a significant loss.

\begin{lemma} \label{conv:1}
    For any $\epsilon > 0$ there exists $\alpha_0 = \alpha_0(\epsilon) > 0$ such that the following happens for all $\alpha < \alpha_0$. For any $\mathbb X \in \mathcal S $ such that $H(\mathbb X ) > 1 + 2\alpha \log \alpha$ and for any $L^*> \log(1/\alpha)$, there
exists  $\mathbb X_{L^*} \in \mathcal S_{L^*} $ such that
\begin{equation}
    I(\mathbb X) \leq I( \mathbb X_{L^*}) + \alpha^{1/2-\epsilon}(L^*)^{-1}\log(L^*).
\end{equation}
\end{lemma}

\begin{proof}
The proof follows a similar structure to that of \cite[Lemma III.2]{kanoria2010deletion} with some modifications.

   We construct $\mathbb{X}_{L^*}$ by flipping a bit each time it is the $(\mathcal{S}_{L^*}+1)$-th consecutive bit with the same value. The density of such bits is upper bounded by $\beta = \frac{P(L_0 > {L^*})}{L^*}$. The corresponding output is denoted by $Y_{L^*} = Y (X^n_{L^*} )$.    
    
    We further define $F = F(\mathbb X, \mathbb A)$ as a binary vector with the same length as the channel output. The elements of $F$ are 1 wherever the corresponding bit in $Y_{L^*}$ is flipped relative to $Y$, and 0 otherwise. The entropy $H(F)$ can be calculated by: 
    %
    \begin{align}
        H(F) &= H(F|L_F) + H(L_F) \\
        &\leq H(F|L_F) + \log(n+1) \\
        &= \sum_{l = n}^{2n} H(F|L_F = l)P(L_F = l) + \log(n+1) \\
        &= \sum_{l = n}^{2n} h(\beta) l P(L_F = l) + \log(n+1) \\
        &= h(\beta) (1+\alpha)n + \log(n+1),
    \end{align}
    where $L_F$ is the length of $F$. The rest of the proof follows the same steps as \cite[Lemma III.2]{kanoria2010deletion}. 
\end{proof}

\begin{lemma}[Converse] \label{conv:2}
    For any $\epsilon > 0$ there exists there exists $ \alpha_0 =  \alpha_0(\epsilon) > 0$ such that the following happens. For any $L^* \in \mathbb N$ and any $\mathbb X \in \mathcal S_{L^*}$ satisfying $H(\mathbb{X}) > 1 - \alpha^{\gamma}$, where $\gamma > \frac{1}{2}${\color{black}{. If $ \alpha <  \alpha_0(\epsilon)$,}} then
    \begin{equation}
        I(\mathbb X ) \leq 1+  \alpha \log  \alpha+ G_1  \alpha +  \alpha^{3/2-\epsilon}. 
    \end{equation}
\end{lemma}

\begin{proof}
With a similar approach to that of~\cite[Lemma III.3]{kanoria2010deletion}, we can write
\begin{equation}
	H(Y) \leq n(1+ \alpha) + \log(n+1),
\end{equation}
since $Y(X^n)$ contains $n+\text{Binomial}(n, \alpha)$ bits.
The proof involves using the lower bound in Corollary \ref{cor1} along with $h(z,v)$.

Note that \( h(z,v) = h(z) + h(v|z) \leq h(z) + h(v) \). Recall the definitions of \(\mathbb{Z}\) and \(\mathbb{V}\) given in \eqref{eqz} and \eqref{eqv}, respectively: \(\mathbb{Z}\) is the process used to keep track of reversed insertions, while \(\mathbb{V}\) stores the inserted bits in the corresponding positions. Note that, due to the relationship between \(\mathbb{Z}\) and \(\mathbb{V}\), we have $h(v) \leq h(z)$ which results in $h(z,v) \leq 2h(z)$.

With these definitions, the expected number of reversed insertions in a run of length \(l\) can be upper bounded by
\begin{equation*}
    l\alpha - l\alpha(1-\alpha)^{l-1} \leq l^2 \alpha^2,
\end{equation*}
where the inequality follows from the fact that \((1-\alpha)^{l-1} \geq 1 - (l-1)\alpha\).

Since a run can have at least length one, we have $z = P(Z_i=1) \leq \alpha^2 \mathbb{E}[L^2]$.

Using \cite[Lemma V.3]{kanoria2013optimal} and ${L^*} \leq 1/\alpha$ with probability 1, we have 
\begin{equation}
    \mathbb{E}[{L}^2] \leq \frac{1}{\alpha} E[L] \leq 20 \alpha^{\gamma-1},
\end{equation}
which implies $  z \leq 20 \alpha^{\gamma+1}$. It follows that 
\begin{equation}
   h({z}) \leq \alpha^{1+\gamma-\epsilon/2},
\end{equation}
for small enough $\alpha$. 

Using $h(z,v)  \leq 2h(z)$, we have
\begin{equation}
   h({z},v) \leq 2\alpha^{1+\gamma-\epsilon/2}.
\end{equation}
Also noting
\begin{equation}
	\big|\mathbb{E}[\log L_0]- \sum_{l=1}^\infty 2^{-l-1} l \log l \big| = o(\alpha^{1/2-\epsilon} \log L^*).
\end{equation}
The proof follows.
\end{proof}

\section{Capacity Approximation for the Gallager Insertion Channel} \label{sec:GalCh}

In this section, we provide the capacity approximation for the Gallager insertion channel model, similar to the technique used for the first insertion channel model. The main result for the capacity is presented in the following theorem.
\begin{theorem}
    Let $C_2(\alpha)$ be the capacity of the Gallager insertion channel with insertion probability $\alpha$. Then, for small $\alpha$ and any $\epsilon > 0$, we have
      \begin{align}
        C_2(\alpha) = 1 + \alpha \log(\alpha)  +G_2 \alpha +  \mathcal{O}(\alpha^{3/2-\epsilon}),
    \end{align}
 where 
 \begin{align}
     G_2 = - &\log(e)  - \frac{7}{8}
     +\frac{1}{4} \sum_{l=1}^{\infty} 2^{-l-1}l \log l \nonumber \\ 
    & + \frac{1}{4}\sum_{a = 1}^\infty \sum_{b = 1}^\infty  (a+b+2)2^{-a}2^{-b}h\left(\frac{a+1}{a+b+2} \right),
 \end{align}
which is approximated as $G_2 \approx  -0.5865$.
\end{theorem}

In the next subsections, we first provide the different terms needed to compute $I(X^n, Y(X^n))$, then show the achievability and converse proofs leading to this approximation.

\subsection{Calculation of $ H(A^n,B^n)$} \label{sec:Hab2}

We follow a similar approach to that used for the simple insertion channel model as described in Sec. \ref{SecHAB}. The insertion pattern process corresponding to the sequence of insertion channels is defined as $\mathbb{A}$, an i.i.d. Bernoulli($\alpha$) process that determines the locations of insertions, independent of the input, and represented by the $n$-bit binary vector $A^n$. Each element of $A^n$, denoted as $A_i$, is set to 1' if the $i$-th bit of the input vector is replaced by two random bits, and 0' otherwise. For the inserted bits, as described in Section \ref{SecHAB}, we use the notation $B_i^j$ for $j \in \{1,2\}$.

\begin{lemma} \label{lemma:Hab2}
We have  
  \begin{align}  
   \lim_{n \rightarrow \infty } \frac{1}{n} H(A^n, B^n)  =  h(\alpha) +2\alpha ,
\end{align}  
where $h(\cdot)$ is the binary entropy function. 
\end{lemma}

\begin{proof}

Note that for this channel model, we have \( B^n = (B^{1,n},B^{2,n}) \). 

The joint probability $P(A_i = a, B^{1,n}_i=b^1, B^{2,n}_i = b^2)$ can be expressed as:
\begin{align*}
       P(a, b^1,  b^2) = 
\begin{cases}
    \frac{\alpha}{4},& \text{if } (a,b^1,b^2) \in \{(1,1,0) ,  (1,0,1) , \\ 
    & \qquad \qquad (1,1,1) , (1,0,0) \} \\
    1-\alpha,              & \text{if } (a,b^1,b^2) = (0,0,0) \\
    0, & \text{otherwise}
\end{cases}
\end{align*}
Utilizing this joint probability, we can derive the following entropy:
\begin{align}
    \frac{1}{n}H(A^n,B^1,B^2) = -\alpha \log & \alpha +2\alpha  -(1-\alpha)\log(1-\alpha).
\end{align}
\end{proof}

\subsection{Calculating $H(A^n, B^n | X^n, Y, K)$} \label{secHabxyk2}
As in Section \ref{sec:Habxyk}, we introduce a modified insertion process for the Gallager insertion channel by defining $\hat{A}^n$ and $\hat{B}^n = (\hat{B}^{1,n}, \hat{B}^{2,n})$. The modified process keeps the runs with at most one insertion as they are while reversing more insertions.

We further define
\begin{equation} \label{eqz2}
    \mathbb{Z}^n = \hat{\mathbb{A}}^n \oplus  \mathbb{A}^n,
\end{equation}
and 
\begin{equation} \label{eqv2}
    \mathbb{V}^n = \hat{\mathbb{B}}^n \oplus  \mathbb{B}^n,
\end{equation}
where $\mathbb{V}^n = (\mathbb{V}^{1,n}, \mathbb{V}^{2,n})$ and can be rewritten as
\begin{align} \label{eqv3}
    \mathbb{V}^{1,n} &= \hat{\mathbb{B}}^{1,n} \oplus  \mathbb{B}^{1,n}, \nonumber \\
    \mathbb{V}^{2,n} &= \hat{\mathbb{B}}^{2,n} \oplus  \mathbb{B}^{2,n}. 
\end{align}

As in the first insertion channel model, we use the following pairs to show that the modified insertion process do not affect the first two order terms:
\begin{align}
    \left( (X^n, Y, K, A^n, B^n), (X^n, \hat Y, \hat{K}, \hat A^n, \hat B^n) \right),
\end{align}
and
\begin{align}
    \left( (X^n, Y, K), (X^n, \hat Y, \hat K ) \right).
\end{align}
With a similar approach, both of these pairs are in the form of $(T,E)$ such that $T$ is a function of $(E, Z^n, V^n)$ and $E$ is a function of $(T, Z^n, V^n)$. Hence, we have:
\begin{equation} \label{hae2}
    |H(T)-H(E)| \leq H(Z^n,V^n),
\end{equation}
where ${V}^n = ({V}^{1,n}, {V}^{2,n})$. As a result, we have:
\begin{align} \label{dif_final2}
    |H(&A^n, B^n | X^n, Y,K) -H(\hat A^n, \hat B^n | X^n,  \hat Y, \hat{K}) |  \leq 2H(Z^n,{V}^{1,n}, {V}^{2,n}).
\end{align}

For these processes and to calculate $H(Z^n,{V}^{1,n}, {V}^{2,n})$, we  have the following probabilities:
\begin{align} \label{probZz2}
     P&(z=a, v^1=b, v^2 = c) = 
\begin{cases}
    & 1- \alpha + \mathbb{E} [\alpha(1-\alpha)^{L_0-1}] , \\ 
    & \qquad \text{if } (a,b,c) = (0,0,0) \\
    & \frac{1}{4} \left(\alpha - \mathbb{E} [\alpha(1-\alpha)^{L_0-1}] \right),             \\
    &  \qquad \text{if } (a,b,c) \in \mathcal{A} \\
    & 0 , \qquad \text{otherwise},  
\end{cases}
\end{align}
where $ \mathcal{A} = \{ (1,0,0) ,(1,0,1),(1,1,0) ,(1,1,1) \}$.

As in Sec. \ref{sec:Habxyk}, one can use \eqref{probZz2} to calculate $H(Z^n,{V}^{1,n}, {V}^{2,n}) \leq n h(z,v^1, v^2)$ and show that it does not affect lower order terms.

\begin{lemma} \label{lemma:habxyk2}
    Consider the Gallager insertion channel model with any $\mathbb X $ such that  $\mathbb{E}[L_0 \log L_0 ] {\color{black}{< \infty}}$. Then, we have
    \begin{align}
        \lim_{n \rightarrow \infty }  \frac{1}{n}H({A}^n,&{B}^n|X^n, {Y}(X^n),K) \nonumber \\
        & =  \frac{1}{4}  \alpha \mathbb{E} [\log (L_0)] 
         + \frac{1}{4}  \alpha \frac{\mathbb{E}[L_0]-1}{\mathbb{E}[L_0]} - \eta,
    \end{align}
    where $ - 2h(z,v^1,v^2)\leq  \eta \leq \frac{1}{4}\alpha^2 \mathbb{E} [L_0 \log (L_0)] + \frac{1}{4} \alpha^2 (\mathbb{E}[L_0]-1)+ 2h(z,v^1,v^2)$.
\end{lemma}

\begin{proof}

The proof follows from \eqref{dif_final2} and Lemma \ref{lemmaABXY_markerG} given in the next part.
\end{proof}

\begin{lemma} \label{lemmaABXY_markerG}
    With Gallager insertion channel model, consider any $\mathbb X $ such that  $\mathbb{E}[L_0 \log L_0 < \infty]$. Then, we have
    \begin{align}
        \lim_{n \rightarrow \infty } \frac{1}{n}  H(\hat{A}^n, & \hat{B}^n  |X^n, \hat{Y}(X^n), \hat K) \nonumber \\ 
       & =  \frac{1}{4}  \alpha \mathbb{E} [\log (L_0)] 
         + \frac{1}{4}  \alpha \frac{\mathbb{E}[L_0]-1}{\mathbb{E}[L_0]}- \delta ,
    \end{align}
    where $ 0\leq  \delta \leq \frac{1}{4}\alpha^2 \mathbb{E} [L_0 \log (L_0)] + \frac{1}{4}  \alpha^2 (\mathbb{E}[L_0]-1)$.
\end{lemma}

\begin{proof}
To simplify understanding of the lemma, let's focus on a run of `0's with length $l_r$, for example:
\begin{equation} \label{run5}
    [{\color{lightgray}{1}} {\color{purple}{\Vert}}  0 \ 0 \ 0 \ 0 \ 0 {\color{purple}{\Vert}} {\color{lightgray}{1}} ]_{1 \times (l_r+2)}.
\end{equation}
For such a run, there are five different cases to consider:

\begin{enumerate}
    \item \textbf{No insertion:} In this case, there is no ambiguity and no contribution to $H(\hat A^n,\hat B^{n}|x^n, \hat y, \hat k)$ as discussed in Section \ref{sec:Habxyk}.
    
    \item \textbf{Replace a bit with `11':} Inserting `11' into the run of `0's reveals all information regarding $\hat D$, $\hat B^{1,n}$, and $\hat B^{2,n}$ given the marker vector $\hat K$. Hence, there'is no contribution to $H(\hat A^n,\hat B^n|x^n, \hat y , \hat k)$.
    
    \item \textbf{Replace a bit with `00':} There are $l_r$ possible locations (each equally likely) for insertion in $\hat D$ given the marker vector $K$. No additional ambiguity arises from $\hat B^n$ (the corresponding bits of $\hat B^{i,n}$ are all zero regardless of the $l_r$ possible insertion positions for $i \in \{1,2\}$). The corresponding contribution is:
    \begin{align} \label{dbxy2_marker}
        &H(\hat A^n,\hat B^n|x^n, \hat y, \hat k)\Big|_{00} = \frac{1}{4} \sum_{i = 1}^n \hat A_i \log(l_{(i)}), 
    \end{align}
    where $l_{(i)}$ is the length of the run containing the $i$-th bit.

Your explanation is clear, but I'll make a few adjustments for presentation and clarity:

\item \textbf{Replace a bit with `01':} For this case, we can consider two disjoint events as follows:
\begin{itemize}
    \item The insertions (replacements with `01') for the first $l_r-1$ bits: Consider the run given in \eqref{run5}. Assume we replace the first bit with `01':
    \begin{equation} \label{run51}
        [{\color{lightgray}{1}} {\color{purple}{\Vert}} {\color{black}  0 \ 1 }\ 0 \ 0 \ 0 \ 0 {\color{purple}{\Vert}} {\color{lightgray}{1}} ]_{1 \times (l_r+2)}.
    \end{equation}
    Now, consider another case in which the second bit is replaced by `10':
    \begin{equation} \label{run52}
        [{\color{lightgray}{1}} {\color{purple}{\Vert}}  0 \ {\color{black} 1 \ 0} \ 0 \ 0 \ 0 {\color{purple}{\Vert}} {\color{lightgray}{1}} ]_{1 \times (l_r+2)}.
    \end{equation}
    Clearly, these two cases result in the same output. Hence, there is a contribution of $\log 2 $ for these first $l_r-1$ bits.
    
    \item The insertion (replacement with `01') for the last bit of the run: If the last bit is replaced with `01', then there will be no ambiguity.
\end{itemize}

We can write the contribution from this case as:
\begin{align} \label{dbxy2_marker01}
     H(\hat A^n,\hat B^n|x^n, \hat y, \hat k)\Big|_{01} &= \frac{1}{4} \sum\limits_{\substack{i=1, \\ i \text{ is not the} \\ \text{last bit of a run}}}^n \hat A_i \log(2). 
\end{align}

\item \textbf{Replace a bit with `10':} Similar to the previous case, we have two disjoint events.
\begin{itemize}
    \item The insertions (replacements with `10') for the last $l_r-1$ bits: Consider the run given in \eqref{run5}. Assume we replace the second bit with `10':
    \begin{equation} \label{run53}
    [{\color{lightgray}{1}} {\color{purple}{\Vert}}  0 \ {\color{black} 1 \ 0} \ 0 \ 0 \ 0 {\color{purple}{\Vert}} {\color{lightgray}{1}} ]_{1 \times (l_r+2)}.
\end{equation}
Now, consider another case in which the first bit is replaced by `01':
\begin{equation} \label{run54}
    [{\color{lightgray}{1}} {\color{purple}{\Vert}} {\color{black}  0 \ 1 }\ 0 \ 0 \ 0 \ 0 {\color{purple}{\Vert}} {\color{lightgray}{1}} ]_{1 \times (l_r+2)}.
\end{equation}
Clearly, these two cases result in the same output. Hence, there is a contribution of $\log 2$ for these last $l_r-1$ bits. We further note that this event intersects the scenario we described previously with the replacement with `01'. Hence, we will discount their intersection while calculating the overall ambiguity. 

\item The insertion (replacement with `10') for the first bit of the run: If the first bit is replaced with `10', then there will be no ambiguity.
\end{itemize}

We can write the contribution from this case as:
\begin{align} \label{dbxy2_marker10}
     H(\hat A^n,\hat B^n|x^n, \hat y, \hat k)\Big|_{01} &= \frac{1}{4} \sum\limits_{\substack{i=1, \\ i \text{ is not the} \\ \text{first bit of a run}}}^n \hat A_i \log(2). 
\end{align}

\end{enumerate}

Hence, considering the insertions with ambiguity (and the intersecting events of insertions with `01' and `10'), the overall contribution to $H(\hat A^n,\hat B^n|x^n, \hat y,\hat k)$ can be written as:
\begin{align} \label{dbxy2_marker_gal}
    H(\hat A^n,\hat B^n|x^n, \hat y, \hat k)= \frac{1}{4}& \sum_{i = 1}^n \hat A_i \log(l_{(i)})   + \frac{1}{4} \sum\limits_{\substack{i=1, \\ i \text{ is not the} \\ \text{first bit of a run}}}^n \hat A_i \log(2). 
\end{align} 
To account for the first bit of the run, \eqref{dbxy2_marker_gal} can be rewritten as:
\begin{align} \label{dbxy2_marker_gal2}
    H(\hat A^n,\hat B^n|x^n, \hat y, \hat k)= \frac{1}{4}  & \sum_{i = 1}^n \hat A_i \log(l_{(i)})  + \frac{1}{4} \sum\limits_{i = 1}^n \hat A_i \log(2) \frac{l_{(i)}-1}{l_{(i)}}. 
\end{align} 
We will use the same definition as in \eqref{pai}:
\begin{equation} \label{pdi2}
    P(\hat A_i = 1)  \in (\alpha-l_{(i)}\alpha^2,\alpha).
\end{equation}
Inserting \eqref{pdi2} into \eqref{dbxy2_marker_gal2}, taking expectation and letting $n \rightarrow \infty$, we have
\begin{align} \label{hatdbxy_marker2x}
    H(&\hat{A}^n,\hat{B}^1,\hat B^2|X^n, \hat{Y}, \hat K)  =  \frac{1}{4} n \alpha \mathbb{E} [\log (L_0)]  + \frac{1}{4} n \alpha \frac{\mathbb{E}[L_0]-1}{\mathbb{E}[L_0]}- \delta n,
\end{align}
where $0\leq  \delta \leq \frac{1}{4}\alpha^2 \mathbb{E} [L_0 \log (L_0)] + \frac{1}{4} \alpha^2 (\mathbb{E}[L_0]-1)$, which completes the proof of Lemma \ref{lemmaABXY_markerG}. 
\end{proof}

\subsection{Calculating $H(K|X^n,Y)$} \label{SecHKXY2}
\begin{lemma}  \label{lemma:kxy2}
    For the Gallager insertion channel {\color{black}{any $\epsilon > 0$, 
there exists $\alpha_0 \equiv \alpha_0(\epsilon) > 0$ such that for any $\alpha < \alpha_0$
and any stationary and ergodic input process $\mathbb{X} \in \mathcal{S}_{\lfloor 1/\alpha \rfloor}$}} satisfying $H(\mathbb{X}) > 1 - \alpha^{\gamma}$, where $\gamma > \frac{1}{2}$, we have:
     \begin{align} 
\lim_{n \rightarrow \infty} \frac{1}{n}  H(K|X^n,Y) = & \frac{\alpha}{4}\sum_{a = 1}^\infty \sum_{b = 1}^\infty  (a +b+2) 2^{-a}2^{-b}   h\left(\frac{a+1}{a+b+2}  \right) + \epsilon_1, 
\end{align}
where
 \begin{equation}
     - \alpha^{1+ \gamma/2-\epsilon }- 4\alpha^{1+ \gamma-\epsilon/2} - \frac{\alpha^2}{4}\sum_{r_j = 1}^\infty  \sum_{r_{j+1} = 1}^\infty  (r_j + r_{j+1}) (r_j +r_{j+1}+2)   h\left(\frac{r_j+1}{r_{j}+r_{j+1}+2}  \right) \leq \epsilon_1 \leq  4\alpha^{1+ \gamma-\epsilon/2} + \alpha^{1+ \gamma/2-\epsilon},
 \end{equation}
 with $\epsilon> 0$. 
\end{lemma}

\begin{proof}
    To compute $H(K|X^n,Y)$, we define the perturbed insertion process $\check{\mathbb{Z}}$ as in Section \ref{SecHKXY}, and follow the same steps. We use the same definitions $\check K$, $\check Y$, and $\bar{Y}$. With this process, we assume that there is at most one insertion on consecutive two runs. 

    We compute $H(\check{K}|X^n,\check{Y})$ and find a bound for $|H({K}|X^n,{Y})- H(\check{K}|X^n,\check{Y})|$, as in the following Lemmas \ref{lem:checkK2} and \ref{lem_Kdif2}, respectively.
\end{proof}

\begin{lemma} \label{lem:checkK2}
For the Gallager insertion channel using the perturbed insertion process with any $\epsilon > 0$, 
{\color{black}{there exists $\alpha_0 \equiv \alpha_0(\epsilon) > 0$ such that for any $\alpha < \alpha_0$}}
and any stationary and ergodic input process $\mathbb{X} \in \mathcal{S}_{\lfloor 1/\alpha \rfloor}$ 
satisfying $H(\mathbb{X}) > 1 - \alpha^{\gamma}$, where $\gamma > \frac{1}{2}$, we have:
    \begin{align} \label{hkxy_last2}
\lim_{n \rightarrow \infty}  H(\check K|X^n,\check Y) = & \frac{\alpha}{4}\sum_{r_j = 1}^\infty \sum_{r_{j+1} = 1}^\infty  (r_j +r_{j+1}+2) 2^{-r_j}2^{-r_{j+1}}   h\left(\frac{r_j+1}{r_{j}+r_{j+1}+2}  \right) - \epsilon_2 + \epsilon_3
\end{align}
which is approximately $\sim 1.4090\alpha - \epsilon_2+\epsilon_3$ where 
\begin{align} \label{eps2}
   0 \leq \epsilon_2 \leq  \frac{\alpha^2}{4}\sum_{r_j = 1}^\infty & \sum_{r_{j+1} = 1}^\infty  (r_j + r_{j+1}) (r_j +r_{j+1}+2)   h\left(\frac{r_j+1}{r_{j}+r_{j+1}+2}  \right),
\end{align}
and 
\begin{equation}
   - \alpha^{1+ \gamma/2-\epsilon} \leq \epsilon_3 \leq \alpha^{1+ \gamma/2-\epsilon}.
\end{equation}
\end{lemma}

    \begin{proof}

    To calculate $H(\check{K}|X^n,\check{Y})$ as in \eqref{hkxy1}, the first step is to calculate
\begin{equation}
    t_j \equiv H\left( |\check Y(j)| \Big|  X(j) \cdots X(M), \bar{Y}(j') \cdots \bar{Y}(M') \right).
\end{equation}

Similar to Lemma \ref{lem:checkK}, we proceed from left to right. Although insertions may introduce new runs under our definition of $\bar{Y}$, the indices of $\check{Y}(j)$ and $\bar{Y}(j')$ may not coincide; however, the corresponding runs still match and represent the same run segment.


We have four cases:
\begin{enumerate}
    \item Case $\mathcal{V}_1$: $|\bar{Y}(j')| = |X(j)|$ and $|\bar{Y}(j'+1)| = |X(j+1)|$: In this case, we certainly do not have insertion in $X(j)$. Hence, $t_j = 0$. 
    
    \item Case $\mathcal{V}_2$: $|\bar{Y}(j')| < |X(j)|$: We have a single bit replacement inside $X(j)$. Consider $X(j)$ as a run of 0's:
    \begin{itemize}
        \item There can be a replacement with `11' or `10' (on any bit location).
        \item There can be a replacement with `01' (except the last bit).
    \end{itemize}
    In those cases, there is no ambiguity in the dimension of $Y(j)$ (or equivalently on $\check K$). Hence, $t_j = 0$.

    \item Case $\mathcal{V}_3$: $|\bar{Y}(j')| = |X(j)|+1$, $|\bar{Y}(j'+1)| = |X(j+1)|$, and there is only a single insertion at the $j$-th and $(j+1)$-th runs (since there will be at most a single insertion in two consecutive runs, we know that there is no insertion at the $(j+2)$-th run).  

    Consider the following example:
            \begin{align}
        X(j)X(j+1) & = [0 \ 0\ 0\ 1\ 1\ 1\ 1],\\
        \bar{Y}(j')\bar{Y}(j'+1) & = [0 \ 0\ 0\ 0\ 1\ 1\ 1\ 1].
    \end{align}
    This can happen in two different scenarios: \\ 
    - If any of the bits in $X(j)$ are replaced by `00', which is with probability:
    \begin{equation}
        P_a = \frac{1}{4} \alpha (1-\alpha)^{|X(j)| +|X(j+1)|-1}|X(j)|.
    \end{equation}
    
    - Or, if the first bit of $X(j+1)$ is replaced by `01', which is with probability:
       \begin{equation}
        P_b = \frac{1}{4} \alpha (1-\alpha)^{|X(j)| +|X(j+1)|-1}.
    \end{equation}
    Note that these two scenarios are disjoint. Hence, we have
    \begin{equation}
        P(\mathcal{V}_3) = \frac{1}{4} \alpha (1-\alpha)^{|X(j)| +|X(j+1)|-1}(|X(j)|+1).
    \end{equation}

        \item Case $\mathcal{V}_4$: $|\bar{Y}(j')| = |X(j)|$, $|\bar{Y}(j'+1)| = |X(j+1)|+1$, and there is only a single insertion at the $j$-th and $(j+1)$-th runs.
        
    Consider the following example:
        \begin{align}
        X(j)X(j+1) & =  [0 \ 0\ 0\ 1\ 1\ 1\ 1],\\
        \bar{Y}(j')\bar{Y}(j'+1) & =  [0 \ 0\ 0\ 1\ 1\ 1\ 1\ 1].
    \end{align}
    This can happen in two different scenarios: \\ 
    - If any of the bits in $X(j+1)$ are replaced by `11', which is with probability:
    \begin{equation}
        P_e = \frac{1}{4}\alpha (1-\alpha)^{|X(j)| +|X(j+1)|-1}|X(j+1)|.
    \end{equation}
    
    - Or, if the last bit of $X(j)$ is replaced by `01', which is with probability:
       \begin{equation}
        P_f = \frac{1}{4} \alpha (1-\alpha)^{|X(j)| +|X(j+1)|-1}.
    \end{equation}
    Using these disjoint events, we have
    \begin{equation}
        P(\mathcal{V}_4) = \frac{1}{4} \alpha (1-\alpha)^{|X(j)| +|X(j+1)|-1}(|X(j+1)|+1).
    \end{equation}

  \end{enumerate}

    Overall, the only ambiguity in $\check K$ comes from two cases: for $|\check{Y}(j)| = |X(j)|$ and $|\check{Y}(j)| = |X(j)|+1$. Hence, by defining $r_j \equiv |X(j)|$, we can write
    \begin{align} \label{yjg}
        |\check{Y}(j)| = 
        \begin{cases}
            r_j + 1,& \text{wp } \frac{1}{4}\alpha(1-\alpha)^{r_j+r_{j+1}-1} (r_j+1) \\
            r_j,     & \text{wp } \frac{1}{4}\alpha(1-\alpha)^{r_j+r_{j+1}-1} (r_{j+1}+1)
        \end{cases}
    \end{align}
Since the only ambiguity on $ |\check{Y}(j)|$ comes from \eqref{yjg}, we can normalize it as:
    \begin{align} \label{yjg2}
        |\check{Y}(j)| = 
        \begin{cases}
            r_j + 1,& \text{wp } \frac{r_j+1}{r_{j}+r_{j+1}+2} \\
            r_j,     & \text{wp } \frac{r_{j+1}+1}{r_{j}+r_{j+1}+2} 
        \end{cases}
    \end{align}
    As a result, we have
    \begin{equation}
        t_j = h\left(\frac{r_j+1}{r_{j}+r_{j+1}+2}  \right).
    \end{equation}
    Thus, the expected contribution of this term to the sum is
    \begin{align}
        &\sum_{r_j = 1}^\infty \sum_{r_{j+1} = 1}^\infty \frac{1}{4} \alpha (1-\alpha )^{r_j+r_{j+1}-1} (r_j +r_{j+1}+2) p_{L(2)} (r_j,r_{j+1}) h\left(\frac{r_j+1}{r_{j}+r_{j+1}+2}  \right),
    \end{align}
where $p_{L(k)}(l_1, \dots, l_k)$ is the joint distribution of $k$ input run lengths.

    Note that we have $(1-\alpha)^{r_j+r_{j+1}-1} \in \left(1-\alpha(r_j+r_{j+1}), 1\right)$, which gives
    \begin{align} \label{kxy_gal}
     \lim_{n \rightarrow \infty} \frac{1}{n}    H(\check K|X^n,Y) = &\frac{\alpha}{4}\sum_{r_j = 1}^\infty \sum_{r_{j+1} = 1}^\infty  (r_j +r_{j+1}+2)  p_{L(2)}(r_j,r_{j+1})  h\left(\frac{r_j+1}{r_{j}+r_{j+1}+2}  \right) - \epsilon_2
    \end{align}
    with 
    \begin{align} \label{eps2_gal}
        0 \leq \epsilon_2 \leq \frac{\alpha^2}{4}\sum_{r_j = 1}^\infty \sum_{r_{j+1} = 1}^\infty  (r_j +r_{j+1}) (r_j +r_{j+1}+2)  h\left(\frac{r_j+1}{r_{j}+r_{j+1}+2}\right).
    \end{align}
    where $p_{L(2)} (r_j,r_{j+1}) \leq 1$ is used in \eqref{eps2_gal}.

   As in Lemma \ref{lem:checkK}, to bound \eqref{kxy_gal}, let us consider \( p^*_{L(k)}(l_1, \dots, l_k) = 2^{- \sum_{i=1}^k l_i} \) and \( H(\mathbb{X}) > 1 - \alpha^\gamma \) with \( \gamma > 1/2 \). Then, by applying \eqref{plk_2} to the first term in \eqref{kxy_gal},
\begin{align} \label{hkxy_last}
\lim_{n \rightarrow \infty} \frac{1}{n} H(\check K|X^n,Y) =  A_1 \alpha - \epsilon_2 + \epsilon_3,
\end{align}
with
\begin{align}
    A_1 = \frac{1}{4}\sum_{r_j = 1}^\infty \sum_{r_{j+1} = 1}^\infty  (r_j +r_{j+1}+2)  2^{-r_j} 2^{-r_{j+1}} h\left(\frac{r_j+1}{r_{j}+r_{j+1}+2}  \right) 
\end{align}
\begin{equation}
   - \alpha^{1+\gamma/2-\epsilon} \leq \epsilon_3 \leq \alpha^{1+\gamma/2-\epsilon},
\end{equation}
and $\epsilon_2$'s range is defined in \eqref{eps2_gal}. 

Note that we have $A_1 \approx 1.4090$.
\end{proof}


\begin{lemma} \label{lem_Kdif2}
With a stationary and ergodic input process $\mathbb{X}$ satisfying $H(\mathbb{X}) > 1 - \alpha^{\gamma}$, where $\gamma > \frac{1}{2}$; the difference between $H(K|X^n,Y)$ and $H(\check{K}|X^n, \check{Y})$, corresponding to the original insertion and perturbed insertion processes with the Gallager insertion channel, respectively, satisfies:
\begin{align}
\lim_{n \rightarrow \infty} \frac{1}{n}  |H(K|X^n,Y) - H(\check{K}|X^n, \check{Y})| \leq 4\alpha^{1+ \gamma-\epsilon/2}.
\end{align}
\end{lemma}
\begin{proof}
    The proof is similar to that of Lemma \ref{lem_Kdif}.
\end{proof}

\subsection{Calculations for $H(Y|X)$}
Using the previously mentioned results in Sections \ref{SecHAB}, \ref{sec:Habxyk} and \ref{SecHKXY}, we obtain the following corollary. 
\begin{corollary} \label{cor2}
For the Gallager insertion channel model with a stationary and ergodic input process $\mathbb{X}$ satisfying $H(\mathbb{X}) > 1 - \alpha^{\gamma}$, where $\gamma > \frac{1}{2}$, we have:
\begin{align}
     \lim_{n \rightarrow \infty } &  \frac{1}{n}  H(Y | X^{n}) \nonumber \\
     & = h(\alpha)  + \alpha \Bigg(2- \frac{1}{4}   \mathbb{E} [\log (L_0)] 
         - \frac{1}{4} \frac{\mathbb{E}[L_0]-1}{\mathbb{E}[L_0]} \nonumber   \\
       &  \qquad -  \frac{\alpha}{4}\sum_{a = 1}^\infty \sum_{b = 1}^\infty  (a +b+2) 2^{-a}2^{-b}  h\left(\frac{a+1}{a+b+2}  \right) \Bigg) \nonumber \\
        &  \qquad + \zeta,
\end{align}
where $ \zeta = \eta + \epsilon_1$ with
\begin{align}
 & - \alpha^{1+ \gamma/2-\epsilon}- 4\alpha^{1+ \gamma -\epsilon/2} - \frac{\alpha^2}{4}\sum_{r_j = 1}^\infty \sum_{r_{j+1} = 1}^\infty (r_j +r_{j+1})  (r_j +r_{j+1}+2)  h\left(\frac{r_j+1}{r_{j}+r_{j+1}+2}\right) \leq \epsilon_1 \nonumber \\
 & \qquad \qquad \qquad \qquad \qquad \qquad \qquad \qquad \qquad \qquad \qquad \qquad \leq  4\alpha^{1+ \gamma-\epsilon/2} + \alpha^{1+ \gamma/2-\epsilon}, \nonumber \\
    &    - 2h(z,v^1,v^2)\leq  \eta \leq \frac{1}{4}\alpha^2 \mathbb{E} [L_0 \log (L_0)] + \frac{1}{4}  \alpha^2 (\mathbb{E}[L_0]-1)+ 2h(z,v^1,v^2), \nonumber
\end{align}
with $\epsilon> 0$. 
\end{corollary}
\begin{proof}
    We have
    \begin{align}
        H(Y | X^{n})       &= H(A^n,B^n) - H(A^n,B^n|X^n, Y,K)  - H(K|X^n,Y).
    \end{align}
    Using Lemmas \ref{lemma:Hab2}, \ref{lemma:habxyk2}, and \ref{lemma:kxy2} completes the proof. 
\end{proof}

\subsection{Achievability and Converse}
The achievability and converse proofs follow a similar structure to those of the first insertion channel model (see Sections \ref{sec:ach1} and \ref{sec:conv1}). 
\begin{lemma}[Achievability]
    Let $\mathbb{X}^*$ be the iid Bernoulli(1/2) process. For any $\epsilon > 0$ with the Gallager insertion channel, we have
     \begin{align}
        C_2(\alpha) = 1 + \alpha \log(\alpha)  +G_2 \alpha +  \mathcal{O}(\alpha^{3/2-\epsilon}).
    \end{align}
\end{lemma}
\begin{proof}
The proof mirrors Section~\ref{sec:ach1}. Since $\mathbb{X}$ is i.i.d. Bernoulli$(1/2)$ of length $n$, and the corresponding output $Y(X^{*,n})$ is also i.i.d. Bernoulli with run length $L_Y = n + \text{Bin}(n, \alpha)$, one can repeat the same steps as in Lemma~\ref{lemma:ach} and show that
\begin{align}
H(Y(X^{*,n})) &=n(1 + \alpha) + \mathcal{O}(\log n) .
\end{align}
Using the estimate of $H(Y|X^{*,n})$ from Corollary~\ref{cor2}, with $P(z, v^1, v^2) = \mathcal{O}(\alpha^2)$ and $\mathbb{E}[L_0 \log L_0] < \infty$, completes the proof.
\end{proof}

For the converse, similar to Lemma~\ref{conv:1}, we start by showing that there is no significant loss if we restrict ourselves to finite-length runs with a maximum length $L^*$.
\begin{lemma}
    For any $\epsilon > 0$ there exists $\alpha_0 = \alpha_0(\epsilon) > 0$ such that the following happens for all $\alpha < \alpha_0$. For any $\mathbb X \in \mathcal S $ such that $H(\mathbb X ) > 1 + 2\alpha \log \alpha$ and for any $L^*> \log(1/\alpha)$, there
exists  $\mathbb X_{L^*} \in \mathcal S_{L^*} $ such that
\begin{equation}
    I(\mathbb X) \leq I( \mathbb X_{L^*}) + \alpha^{1/2-\epsilon}(L^*)^{-1}\log(L^*).
\end{equation}
\end{lemma}

\begin{proof}
The proof follows a similar structure to that of \cite[Lemma III.2]{kanoria2010deletion} and Lemma~\ref{conv:1}. Since the steps are provided in detail in Lemma~\ref{conv:1}, we omit them here and instead provide a sketch of the proof.

We construct $\mathbb{X}_{L^*}$ by flipping a bit each time it is the $(\mathcal{S}_{L^*}+1)$-th consecutive bit with the same value, with the corresponding output denoted by $Y_{L^*} = Y(X^n_{L^*})$.

We further define $F = F(\mathbb{X}, \mathbb{A})$ as a binary vector of the same length as the channel output. The elements of $F$ are set to 1 wherever the corresponding bit in $Y_{L^*}$ is flipped relative to $Y$, and 0 otherwise. The entropy $H(F)$ can be computed as:
\begin{align}
    H(F) = h(\beta)(1 + \alpha)n + \log(n + 1),
\end{align}
where $L_F$ is the length of $F$ and $\beta = \frac{P(L_0 > L^*)}{L^*}$. The remainder of the proof follows the same steps as in \cite[Lemma III.2]{kanoria2010deletion}.
\end{proof}

\begin{lemma}[Converse]
    For any $\epsilon > 0$ there exists there exists $ \alpha_0 =  \alpha_0(\epsilon) > 0$ such that the following happens. For any $L^* \in \mathbb N$ and any $\mathbb X \in \mathcal S_{L^*}$ satisfying $H(\mathbb{X}) > 1 - \alpha^{\gamma}$, where $\gamma > \frac{1}{2}${\color{black}{. If $ \alpha <  \alpha_0(\epsilon)$,}} then
    \begin{equation}
        I(\mathbb X ) \leq 1+  \alpha \log  \alpha+ G_2  \alpha + \alpha^{3/2-\epsilon}.
    \end{equation}
\end{lemma}

\begin{proof}
With a similar approach to that of~\cite[Lemma III.3]{kanoria2010deletion} and our proof of Lemma~\ref{conv:2}, we can directly write
\begin{equation}
    H(Y) \leq n(1 + \alpha) + \log(n + 1).
\end{equation}
The proof involves using the lower bound in Corollary~\ref{cor2} together with the function $h(z, v^1, v^2)$. Note that $h(z, v^1, v^2) \leq h(z) + h(v^1) + h(v^2)$. Owing to the similarity between the definitions of $\mathbb{Z}$, $\mathbb{V}^1$, and $\mathbb{V}^2$ and those of $\mathbb{Z}$ and $\mathbb{V}$ in the simple insertion channel, one can readily show that $h(z, v^1, v^2) \leq 3h(z)$. Hence, we have
\begin{equation}
   h({z},v_1,v_2) \leq 3\alpha^{1+\gamma-\epsilon/2}.
\end{equation}
Also noting that
\begin{equation}
    \left| \mathbb{E}[\log L_0] - \sum_{l=1}^\infty 2^{-l-1} l \log l \right| = o(\alpha^{1/2 - \epsilon} \log L^*),
\end{equation}
the proof follows.
\end{proof}

\section{Conclusions} \label{sec:disc}


In this paper, we focused on the capacity of insertion channels with small insertion probabilities, a topic that has been largely overlooked in the literature despite its relevance in practical applications such as DNA coding and storage. Our approach builds upon the series expansion of a previously known capacity point, specifically a channel with no insertions (i.e., zero insertion probability), where the capacity is 1 and is achieved with an i.i.d. Bernoulli$(1/2)$ input distribution. By leveraging this known point, we expect that for small insertion probabilities, the achievable rate will not deviate significantly from this value. With this intuition, we derived the first two terms of the capacity expansion by decomposing the expressions necessary to achieve the maximum rate. Using this methodology, we obtained capacity approximations for two fundamental insertion channel models: 1) the insertion channel with random bit insertions, and 2) the Gallager insertion channel, where a bit is replaced by two random bits. Our results demonstrate that these approximations are achievable by a uniform i.i.d. input distribution up to the first two orders, as expected.  
For future work, possible extensions include applying similar methods to different channel models or symbol alphabets. For instance, the binary input case could be extended to nonbinary inputs, e.g., 4-ary for DNA storage systems, with deletion and/or insertion channels. Further research could also consider different channel models, such as the absorption channel or the insertion, deletion, and substitution (IDS) channel. Thus, our approach lays the foundation for various extensions and can facilitate capacity analysis for these different models.

\bibliographystyle{IEEEtran}
\bibliography{bibs_insertion}

\end{document}